\documentclass[12pt]{iopart}

\usepackage{iopams}  
\usepackage{graphicx}
\usepackage{color}
\begin{document}

\title[]{Turbulent momentum transport due to neoclassical flows}

\author{Jungpyo Lee}
\address{Plasma Science and Fusion Center, Massachusetts Institute of Technology, MA, USA}
\ead{Jungpyo@psfc.mit.edu}

\author{Michael Barnes}
\address{Rudolf Peierls Centre for Theoretical Physics, University of Oxford, UK}
\address{Culham Centre for Fusion Energy, Culham Science Centre, Abingdon, UK}
\author{Felix I Parra}
\address{Rudolf Peierls Centre for Theoretical Physics, University of Oxford, UK}
\address{Culham Centre for Fusion Energy, Culham Science Centre, Abingdon, UK}

\author{Emily Belli}
\address{ General Atomics, CA, USA}
\author{Jeff Candy}
\address{ General Atomics, CA, USA}

\date{\today}


\begin{abstract}
 Intrinsic toroidal rotation in a tokamak can be driven by turbulent momentum transport due to neoclassical flow effects breaking a symmetry of turbulence. In this paper we categorize the contributions due to neoclassical effects to the turbulent momentum transport, and evaluate each contribution using gyrokinetic simulations. We find that the relative importance of each contribution changes with collisionality. For low collisionality, the dominant contributions come from neoclassical particle and parallel flows. For moderate collisionality, there are  non-negligible contributions due to neoclassical poloidal electric field and poloidal gradients of density and temperature, which are not important for low collisionality. 
\end{abstract}

\maketitle

\section{Introduction}
Significant ion toroidal flow has been observed in many tokamaks without any external momentum injection \cite{Rice:NF2007}. The origin of this intrinsic rotation is primarily turbulent momentum redistribution. Understanding the intrinsic rotation observed in experiments is useful because the toroidal flow stabilizes MHD modes \cite{Bondeson:PRL1994, Strait:PRL1995} and increases the confinement time \cite{Burrell:POP1999,Barnes:PRL2011shear,Highcock:PRL2010,Parra:PRL2011}. In this paper, we investigate the radial turbulent momentum transport due to neoclassical flows, which always exist in tokamak plasmas. 
In \cite{Parra:2014} it is argued that this type of momentum redistribution dominates over other momentum flux drives when the size of the turbulent eddies is sufficiently small. Comparisons with experiments \cite{Hillesheim:2014} indicate that this type of momentum flux may explain certain features of intrinsic rotation, such as rotation reversals \cite{Bortolon:PRL2006}.

There are momentum sources and sinks at the boundary of the confinement region, and these sources and sinks likely determine the rotation at the separatrix. The radial transport of toroidal angular momentum redistributes the momentum within a tokamak, and determines the size and sign of the ion toroidal flow given the boundary condition. 
If there is no external momentum injection, the change of the ion rotation is determined by the total momentum conservation equation
\begin{eqnarray}
\frac{\partial (n_i m_i \mathbf{V_i})}{\partial t}=-\nabla \cdot \left[\stackrel{\leftrightarrow}{\bold{p_i}}+p_{e\perp}(\stackrel{\leftrightarrow}{\bold{I}}-\hat{\mathbf{b}}\hat{\mathbf{b}})+p_{e\|}\hat{\mathbf{b}}\hat{\mathbf{b}}\right]+\frac{1}{c} \mathbf{J}\times \mathbf{B},\label{rotation0}
\end{eqnarray}
where $n_i$, $m_i$, and $\mathbf{V_i}$ are the ion density, mass, and fluid velocity, $\stackrel{\leftrightarrow}{\bold{p_i}}=\int d^3 v f_i m_i \mathbf{v}\mathbf{v}$ is the ion pressure tensor, $f_i$ is the ion distribution function, $\hat{\mathbf{b}}$ is the unit vector parallel to the magnetic field $\mathbf{B}$, $p_{e\bot}$ and $p_{e\|}$ are the electron pressure perpendicular and parallel to the magnetic field, $\stackrel{\leftrightarrow}{\bold{I}}$ is the identity tensor, $c$ is the speed of light, and $\mathbf{J}$ is the current density. Here, the small electron inertial term and the small electron viscosity terms are neglected. Multiplying the toroidal component of equation (\ref{rotation0}) by major radius $R$ and taking the flux surface average $\langle ...\rangle_\psi= (1/V')\int  d\theta d\varphi (.../\mathbf{B} \cdot \nabla \theta)$ and the average over several turbulence characteristic times and length scale $\langle ...\rangle_T$ result in the equation
 \begin{eqnarray}
\frac{\partial }{\partial t}\left \langle\left \langle n_im_i R V_{\varphi} \right \rangle_\psi\right\rangle_T= -\frac{1}{V'} \frac{\partial }{\partial \psi}(V'\Pi)+\frac{1}{c} \left \langle\left \langle \mathbf{J} \cdot \nabla \psi \right \rangle_\psi\right\rangle_T\label{rotation1},
\end{eqnarray}
where $V'=\int d\theta d\varphi (\mathbf{B} \cdot \nabla \theta)^{-1}$ is the flux surface differential volume, and $\Pi = m_i \langle\left \langle \int d^3v f_i R(\mathbf{v} \cdot \hat{\boldsymbol{\varphi}}) (\mathbf{v}\cdot \nabla \psi)\right \rangle_\psi\rangle_T$ is the radial transport of ion toroidal angular momentum. Here, $\theta$ and $\varphi$ are the poloidal and toroidal angle, respectively, and $\psi$ is the poloidal magnetic flux, giving the magnetic field $\mathbf{B}= I \nabla \varphi +\nabla \varphi \times \nabla \psi$, where $I=B_\varphi R$ and $B_\varphi$ is the toroidal magnetic field. The transport average $\langle ...\rangle_T$  is used for averaging over turbulent phenomena that have much shorter time and length scales than those in which transport happens. The radial current in the second term on the right hand side vanishes due to quasineutrality, so the change of ion toroidal angular momentum is determined by radial momentum transport,
 \begin{eqnarray}
\frac{\partial }{\partial t}\left \langle\left \langle n_im_i R V_{\varphi} \right \rangle_\psi \right \rangle_T= -\frac{1}{V'} \frac{\partial }{\partial \psi}(V'\Pi)\label{rotation2}.
\end{eqnarray}
The radial transport of particle, energy and momentum in a tokamak is caused dominantly by microturbulence (turbulent transport) rather than by collisional processes (classical and neoclassical transport). Microturbulence in a tokamak is generated mainly by drift wave instabilities due to radial gradients of temperature and density. The radial momentum transport is more complex than particle and energy transport because a symmetry of the turbulence in an up-down symmetric tokamak makes the radial momentum transport vanish to lowest order in an expansion in $\rho_\star$, where $ \rho_{\star}\equiv\rho_i/a\ll1 $, $\rho_i$ is the ion Larmor radius and $a$ is the minor radius \cite{Sugama:PPCF2011,parra2011up} (see section 2). The symmetry of the turbulence is broken by flow and flow shear. The effect of the flow and its gradient on the momentum transport can be linearized in the low flow (sub-sonic flow) regime, 
giving an advective term and a diffusive term that are proportional to the flow and the flow shear, respectively,
\begin{eqnarray}
\Pi\equiv\Pi_{{int}}-P_{\varphi} n_i m_i \langle R^2\rangle_\psi \Omega_{\varphi}-\chi_{\varphi} n_i m_i\langle R^2 \rangle_\psi\frac{\partial\Omega_{\varphi}}{\partial r},\label{Pi_com1}
\end{eqnarray}
where $\Omega_{\varphi}(r)$ is the ion toroidal angular frequency, and $r=a({\psi}/{\psi_0})$ is the radial coordinate that labels flux surfaces. Here, $a$ and  $\psi_0$ are the minor radius and the poloidal flux that label the last closed flux surface. 
The advective term has the momentum pinch coefficient $P_{\varphi}$, and the diffusive term is proportional to the momentum diffusivity $\chi_{\varphi}$. Here, $\Pi_{{int}}$ is the intrinsic momentum transport, also called the residual stress, which is the momentum flux generated even for zero flow and flow shear (i.e. $\Omega_{\varphi}=0$ and ${\partial \Omega_{\varphi}}/{\partial r}=0$). In steady state, the total toroidal angular momentum transport should be zero if there is no external source (i.e. $\Pi=0$ in equation (\ref{Pi_com1})), and the balance between non-zero pieces in $\Pi$ determines the radial profiles of the toroidal flow, $\Omega_{\varphi}(r)$. 

The momentum advection and diffusion have been theoretically investigated in previous work \cite{Peeters:PRL2007,Hahm:POP2007,Peeters:POP2009,Peeters:NF2011}, and theory has been compared with experimental observations in many tokamaks with strong E$\times$B flow (Mach number $\sim$ 1) driven by strong neutral beams \cite{scott:PRL1990, Lao_shear, Menard, solomon2008momentum}. The momentum advection is not just due to the momentum carried out by the particle transport, but also due to inherent inward momentum transport (momentum pinch). The momentum pinch is driven by the drift due to the Coriolis force in the rotating frame \cite{Peeters:POP2009}. The flow shear is diffused out  by the same mechanism by which the ion thermal energy is transported. Consequently, the diffusion rates for the momentum and the ion thermal energy are similar, and the Prandtl number (the ratio of the momentum diffusivity to the ion heat diffusivity) is found to be 0.5-0.8 in many gyrokinetic simulations \cite{Barnes:PRL2011shear,Highcock:PRL2010,Peeters:NF2011,casson:POP2009}. We have found that this previous analysis for $P_{\varphi}$ and $\chi_{\varphi}$ corresponds only to the E$\times$B flow, and it is not valid for the diamagnetic component of the neoclassical flow \cite{Barnes:PRL2013,Lee:NF2014}. 


The intrinsic momentum transport accounts for all other momentum transport mechanisms that are not the pinch or diffusion. This intrinsic momentum transport determines the intrinsic rotation. Many features of experimental profiles of rotation (e.g. the sign change of the rotation at mid-radius in some discharges) cannot be explained by only momentum pinch and diffusion, and so require intrinsic momentum redistribution. The intrinsic momentum transport must be driven by the mechanisms that break the symmetry of the turbulent momentum flux in a non-rotating state (e.g. up-down asymmetric magnetic equilibrium \cite{Camenen:PRL2010,Ball:PPCF2014}, the slow radial variation of plasma parameters \cite{camenen2011consequences,waltz2011gyrokinetic} and the slow poloidal variation of turbulence  \cite{sung2013toroidal}, neoclassical flows \cite{Lee:NF2014,Parra:PPCF2010,Parra:NF2011,Lee:PoP2014}, finite orbit widths \cite{Parra:PPCF2010,Parra:NF2011}). Reference  \cite{Parra:2014} treats all these symmetry breaking effects in a self-consistent way.

In this paper, we focus on the intrinsic turbulent momentum transport due to neoclassical phenomena. We find that the diamagnetic component of the neoclassical parallel particle flow and the E$\times$B flow break the symmetry of the turbulence differently \cite{Lee:NF2014}. As a result, it is possible to obtain turbulent momentum transport for non-rotating states in which the diamagnetic flow and the E$\times$B flow cancel each other. Moreover, the neoclassical parallel heat flow also results in significant toroidal momentum transport. Because the size of the diamagnetic flow (Mach number $\lesssim$ 0.1) is smaller than the ion thermal velocity by $(B/B_\theta)\rho_i/L_T$, calculating the effect of the diamagnetic flow requires higher order corrections to the gyrokinetic equations \cite{Barnes:PRL2013,Lee:NF2014,Parra:PRL2012}. Here, $B$ and $B_\theta$ are the magnitude of total and poloidal magnetic field, respectively, and $L_T$ is the characteristic length of the temperature gradient. The momentum flux due to diamagnetic effects may explain the collisional dependence of the observed intrinsic rotation \cite{Hillesheim:2014,Barnes:PRL2013,Lee:PoP2014} and the rotation peaking during L-H transitions \cite{Lee:NF2014}. We investigate the effect of each piece of the diamagnetic corrections separately in section 4.

The rest of this paper is organized as follows: In section 2, we introduce gyrokinetics and review the symmetry of the turbulence that results from the lowest order gyrokinetic analysis. In section 3, the neoclassical particle and heat flows are discussed. They are used to correct the lowest order gyrokinetic equation. In section 4, using higher order corrections, we evaluate the intrinsic turbulent momentum transport due to neoclassical effects for low and moderate collisionality. We show the momentum flux has contributions from (1) the neoclassical particle flow, (2) the neoclassical parallel heat flow, and (3) the neoclassical poloidal electric field and poloidal gradients of density and temperature. Finally, a discussion is given in section 5.

\section{Gyrokinetics and Symmetry}\label{GKE}
Gyrokinetics is a kinetic description of electromagnetic (or electrostatic) turbulence that averages over the gyromotion while keeping finite Larmor radius effects \cite{catto1981generalized,Frieman:PF1982}. The frequency of drift-wave microturbulence is much smaller than the gyrofrequency, permitting the separation of the gyromotion from the parallel streaming and the perpendicular drift motion. Particle motion is described by the position of the gyration center, and only two velocity space variables (for example, kinetic energy $\mathcal{E}={v^2}/{2}$ and magnetic moment $\mu={v_\perp^2}/{2B}$). Here, $v_\perp=\sqrt{v^2-v_\|^2}$ and  $v_\|$ are the velocities perpendicular and parallel to the static magnetic field, $\mathbf{B}$, respectively. Gyrokinetics assumes $\rho_i/l_\perp \sim 1$  and ${l_\perp}/l_\| \sim \rho_i/a \equiv \rho_{\star} \ll 1$. Here, $l_\|$ and $l_\perp$ are the turbulence length scales parallel and perpendicular to the static magnetic field, respectively. For simplicity, we will only use gyrokinetic equations for electrostatic turbulence which is the most relevant in low $\beta$ tokamaks.

The radial flux of ion toroidal angular momentum in equation (\ref{rotation1}) is dominantly due to the fluctuating radial E$\times$B drift carrying toroidal angular momentum, and it can be evaluated using
\begin{eqnarray}
\Pi\simeq \Pi^{tb}=\left\langle m_i\int d^3 v f^{tb}_i (\mathbf{v}\cdot {\hat{\boldsymbol{\varphi}}}R) (\mathbf{v}^{tb}_{E}\cdot \nabla{\psi})  \right\rangle_\psi\label{Pi_def}, 
\end{eqnarray}
where $\mathbf{v}^{tb}_{E}\cdot \nabla{\psi}$ is the radial drift due to the electrostatic turbulence and $f_i^{tb}$ is the turbulent piece of the ion distribution function. The fluctuating radial E$\times$B drift and distribution function are given by the gyrokinetic equations. The radial flux of the toroidal angular momentum can be also decomposed into a piece due to the fluctuation of the parallel velocity and a piece due to the fluctuation in the perpendicular velocity, $\Pi^{tb}=\Pi_{\|}+\Pi_{\perp}$. The parallel contribution to $\Pi^{tb}$ in equation (\ref{Pi_def}) is 
\begin{eqnarray}
\Pi_\|=\left\langle m_i\int d^3 v f^{tb}_i \left(v_\| \frac{I}{B}\right) (\mathbf{v}^{tb}_{E}\cdot \nabla{\psi})  \right\rangle_\psi\ ,\label{pi_par}
 \end{eqnarray} 
and it dominates over the perpendicular contribution $\Pi_\perp\sim ({B_\theta}/{B})\Pi_\|$ when ${B_\theta}/{B} \ll 1$. In most tokamaks, the poloidal magnetic field $B_\theta$ is much smaller than the total magnetic field $B$.

 The radial momentum transport can be expanded in the small parameter  $\rho_{\star,\theta}\equiv(B/B_\theta)\rho_\star$ that measures the radial drift orbit width with respect to the size of the tokamak, i.e. $\Pi^{tb}=\Pi_1+\Pi_2+...$. The lowest order radial flux ($\Pi_1$) corresponds to the size of the gyro-Bohm diffusion of thermal velocity, $\Pi_1\sim D_{gB,i} ({n_im_i v_{ti}R}/{a})= \rho_\star^2 n_im_iv_{ti}^2R$, where $D_{gB,i}= \rho_\star \rho_iv_{ti} $ is the gyro-Bohm diffusion coefficient. The next order radial flux ($\Pi_2\sim( B/B_\theta)\rho_\star \Pi_1$) is given by the higher order gyrokinetics described in section \ref{sec:high_GKE}. The lowest order gyrokinetic equations described in this section can be used to evaluate the gyro-Bohm scale transport of particles and energy, but the lowest order radial flux of momentum ($\Pi_1$) is different from zero only in the presence of sonic plasma flow, otherwise $\Pi_1=0$ as we will prove. 
 
 The distribution function of species $s$ is assumed to have a non-fluctuating background piece $f^{bg}_s$ and a fluctuating piece due to turbulence $f^{tb}_s$ (i.e. $f_s=f^{bg}_s+f^{tb}_{s}$).  To lowest order in $\rho_\star$, the distribution function is approximately $f_s\simeq f_{0,s}^{bg}+f^{tb}_{1,s}$. The subscript denotes the order in $\rho_\star$. The plasma is sufficiently collisional that the background piece is $f_{0,s}^{bg}= f_{M,s}$, where $f_{M,s}$ is a stationary Maxwellian. The size of the fluctuating piece is much smaller than the background piece, $f^{tb}_{1,s}\sim \rho_{\star}f_{M,s}$. The electrostatic potential can also be split into a background and a fluctuating piece $\phi=\phi^{bg}_0+\phi^{tb}_1$, where $\phi^{bg}_0$ is the background potential, of order $T_e/e$, $\phi^{tb}_1$ is the short wavelength turbulent potential whose size is $O(\rho_{\star}({T_e}/{e}))$, $T_e$ is the electron temperature and $e$ is the electron charge. 
The lowest order gyrokinetic equation for $f^{tb}_{1,s}$ is
\begin{eqnarray}
&\frac{\partial f^{tb}_{1,s}}{\partial t} - \frac{c}{B} \left(\nabla\phi^{bg}_0 (\psi_0) \times \mathbf{\hat{b}}\right) \cdot  \nabla f^{tb}_{1,s}+\left(v_\| \mathbf{\hat{b}}+\mathbf{v}_{M,s}- \frac{c}{B} \nabla\langle\phi^{tb}_1 \rangle \times \mathbf{\hat{b}}\right) \cdot  \nabla f^{tb}_{1,s}\nonumber \\&=\frac{c}{B} \nabla\langle\phi^{tb}_1 \rangle \times \mathbf{\hat{b}} \cdot  \nabla f_{0,s}^{bg} + \frac{Z_se}{m_s}\left(  v_\| \mathbf{\hat{b}}+\mathbf{v}_{M,s}\right) \cdot\nabla\langle\phi^{tb}_1 \rangle\frac{\partial f_{0,s}^{bg}}{\partial \mathcal{E}}+ \langle C(f_s) \rangle \label{GKE1},
\end{eqnarray} 
where $C$ is the collision operator and $\mathbf{v}_{M,s} = ({\mu}/{\Omega_s})\mathbf{\hat{b}} \times \nabla B+({v_\|^2}/{\Omega_s})\mathbf{\hat{b}} \times (\mathbf{\hat{b}} \cdot \mathbf{\nabla \hat{b}})$ is the sum of the $\nabla B$ and curvature drift. Here, ${m_s}$ is the mass and $Z_se$ is the charge of the species, $\Omega_s={Z_seB}/{m_s}$ is the gyrofrequency, and $\langle ...\rangle$ is the average over the gyromotion. Note that $\phi^{bg}_0$ is evaluated at the flux surface of interest $\psi=\psi_0$ in the second term on left hand side of this equation (\ref{GKE1}). We ignore the dependence of $\phi_0^{bg}$ on $\psi$ to this order.

The gyrokinetic equations for different species are coupled by imposing the quasineutrality condition,
\begin{eqnarray}
\sum_s Z_s e \int d^3v \left(f^{tb}_{1,s} -\frac{Z_s e(\phi_1^{tb}-\langle\phi_1^{tb}\rangle)}{T_s}f_{M,s}\right)=0 \label{QN},
\end{eqnarray}
where the second term in the parentheses on the left hand side is the polarization density due to the gyromotion.

Because the turbulence parallel length scales are much larger than the perpendicular length scales ($l_\perp/l_\| \sim \rho_\star$), it is convenient to use the field-aligned coordinates ($\psi$, $\alpha$, and $\theta$) to solve equation (\ref{GKE1}), with the poloidal angle $\theta$ measuring the distance along $B$. The coordinates $\psi$ and $\alpha$ are perpendicular to the magnetic field. The angle $\alpha$ is 
\begin{eqnarray}
\alpha=\varphi-I(\psi)\int_0^{\theta} d\theta^\prime \frac{1}{R^2 \mathbf{B}\cdot \nabla \theta^\prime} \bigg|_{\psi,\theta^\prime}.\label{alpha}
\end{eqnarray}
The static magnetic field satisfies $\mathbf{B}=\nabla \alpha \times \nabla \psi$. 
The slow variation in the parallel coordinate ($\theta$) and the fast variation in the perpendicular coordinates ($\psi$ and $\alpha$) of the turbulence are well described by solutions of the form
\begin{eqnarray}
f^{tb}_{1,s}(\psi,\alpha,\theta; \mathcal{E},\mu,\sigma; t)&=&\sum_{k_\psi,k_\alpha}\hat{f^{tb}_{1,s}}( k_\psi,k_\alpha,\theta; \mathcal{E},\mu,\sigma; t) e^{i(k_\psi \psi +k_\alpha \alpha)},\label{form_f1}\\
\phi_1^{tb}(\psi,\alpha,\theta; t)&=&\sum_{k_\psi,k_\alpha}\hat{\phi}^{tb}_1(k_\psi,k_\alpha,\theta; t) e^{i(k_\psi \psi +k_\alpha \alpha)},\label{form_f2}
\end{eqnarray}
where $k_\psi$ and $k_\alpha$ are the wavevectors in $\psi$ and $\alpha$ coordinates, respectively, and $\sigma$ is the sign of the parallel velocity ($v_\|=\sigma \sqrt{2(\mathcal{E}-\mu B)}$). 


For a poloidally up-down symmetric tokamak, the lowest order gyrokinetic equations (\ref{GKE1}) and  (\ref{QN}) satisfy the symmetry \cite{Sugama:PPCF2011,parra2011up}
\begin{eqnarray} 
\hat{f^{tb}_{1,s}}(\theta,v_\|,k_\psi)= -\hat{f^{tb}_{1,s}}(-\theta,-v_\|,-k_\psi)\label{f_tb1_sym}, \\ \hat{\phi}_1^{tb} (\theta,k_\psi)= -\hat{\phi}_1^{tb} (-\theta,-k_\psi)\label{phi_tb1_sym}. 
\end{eqnarray}

To see the effect of this symmetry on the transport of ion toroidal angular momentum, the parallel contribution in (\ref{pi_par}) is written as an integral and a summation,
\begin{eqnarray}
 \Pi_\|&\equiv&\sum_{k_\psi}\int_{-\infty}^{\infty} d\theta dv_\| \pi_\|(\theta,v_\|,k_\psi)\label{Pi_def0},
 \end{eqnarray}
 where 
 \begin{eqnarray}
 \pi_\|=\frac{2\pi m_i I}{\mathbf{B}\cdot \nabla\theta}\frac{2\pi}{V'}\sum_{k_\alpha}\int_0^{\infty} d\mu (\hat{f^{tb}_i})^\ast   v_\| ik_\alpha \hat{\phi^{tb}} J_0(k_\perp v_\perp /\Omega_i) .
  \end{eqnarray} 
  The asterisk on $f^{tb}_s$ indicates complex conjugation, $J_0$ is a Bessel function of the first kind, and $k_\perp=\sqrt{k_\psi^2 |\nabla \psi|^2+k_\alpha^2 |\nabla \alpha|^2}$.
  The parallel contribution $\Pi_\|$ and the integrand $\pi_\|$ are expanded in  $\rho_{\star,\theta}$ (i.e. $\Pi_\|=\Pi_{\|1}+\Pi_{\|2}+...$). Accordingly, $\Pi_{\|1}=\sum_{k_\psi}\int_{-\infty}^{\infty} d\theta dv_\| \pi_{\|1}$ is the lowest order radial flux and $\Pi_{\|2}=\sum_{k_\psi}\int_{-\infty}^{\infty} d\theta dv_\| \pi_{\|2}$ is the higher order flux in $\rho_{\star,\theta}$. 

From the symmetry properties in equations (\ref{f_tb1_sym},\ref{phi_tb1_sym}), the momentum transport is antisymmetric under the inversion of the poloidal angle, the parallel velocity, and the radial wave vector,
\begin{eqnarray}
\pi_{\|1}(\theta,v_\|,k_\psi)= -\pi_{\|1} (-\theta,-v_\|,-k_\psi).\label{symm_pi1}
\end{eqnarray}
The perpendicular contribution $\Pi_\perp$ has the same symmetry as the parallel contribution $\Pi_\|$ has \cite{Sugama:PPCF2011,parra2011up}. As a result, the turbulent momentum transport vanishes (i.e. $\Pi_1=0$) due to the symmetry of the gyrokinetic solutions unless there is a symmetry breaking mechanism.


\section{Symmetry breaking}\label{sec:sym_break}
There are several effects that break the symmetry of the lowest order gyrokinetic equations described in the previous section. If the toroidal rotation is large (Mach number $\sim$ 1), the symmetry of the gyrokinetic equation is broken even at lowest order (i.e. the symmetry in (\ref{symm_pi1}) is not valid because $\phi_0^{bg}$ is not $O(T/e)$, but $O(\rho_\star^{-1}T/e)$). However, we are interested in describing intrinsic rotation, which is typically small (Mach number $\sim$ 0.1). The symmetry breaking mechanisms for the intrinsic toroidal rotation are described below.  

The symmetry in $\theta$ is broken in up-down asymmetric tokamaks \cite{Camenen:PRL2010}. Typically the up-down asymmetry is significant only around the edge. Some dedicated experiments in the TCV tokamak \cite{Camenen:PRL2010} with significant up-down asymmetry in the core found rotation of about 3 $\%$ of the ion thermal speed. A theoretical study in \cite{Ball:PPCF2014} estimated rotation due to the asymmetry to be of the order of $5\%$ of the ion thermal speed. 

The slow variation of plasma parameters in the radial direction breaks the symmetry in $k_\psi$ \cite{camenen2011consequences,waltz2011gyrokinetic,wang2010}. The effect on the turbulence of the change of temperature and density gradients across the radial dimension of plasma eddies has been investigated in gyrokinetic global codes. For a sufficiently small radial correlation length of the turbulence, the effect of the slow radial variation of the gradients is too small to break the symmetry significantly. 

The higher order correction to the gyrokinetic equation due to the slow poloidal variation of the turbulence can break the symmetry in the poloidal angle $\theta$ \cite{sung2013toroidal}. This effect will be small if the parallel correlation length of the eddies is small compared to the connection length $qR$ \cite{Parra:2014}, with $q$ the safety factor. 

In this paper, we focus on the effect of the neoclassical flow that introduces a preferential direction in the system \cite{Barnes:PRL2013,Lee:NF2014,Lee:PoP2014,Parra:PRL2012}. For turbulence in which the radial and parallel correlation lengths of the turbulence are sufficiently small, these diamagnetic flows dominate over the momentum driven by the slow radial and poloidal variations of the turbulence \cite{Parra:2014}. 

The toroidal flow in a tokamak is composed of two different types of flow: the E$\times$B flow and the neoclassical or diamagnetic flow. 
The diamagnetic flow due to the radial pressure gradient always exists in a tokamak regardless of the size and the sign of the radial electric field. This implies that there is an inherent preferential direction that breaks the symmetry. We call these flows diamagnetic flows to emphasize that they are the result of finite orbit width effects, as shown in figure 1. 
The higher flux for the ions with positive parallel velocity is due to the radial magnetic drift to inner radii where the plasma has a higher temperature and density than at outer radii. Consequently, the amount of diamagnetic flow is determined by the width of the deviation of the poloidal orbit from the flux surface ($\sim ({B}/{B_\theta}) \rho_i$) and the pressure gradient (i.e. $\Omega_{\varphi,d} R \sim \rho_{\star,\theta} v_{ti}$, where $\Omega_{\varphi,d}$ is the toroidal angular frequency for diamagnetic flow). To take into account the diamagnetic flow in gyrokinetics, one needs to correct the lowest order gyrokinetic equation in (\ref{GKE1}) with the higher order terms in $\rho_{\star,\theta}$ as will be explained in section \ref{sec:high_GKE}. 

\begin{figure} 
\includegraphics[scale=0.5]{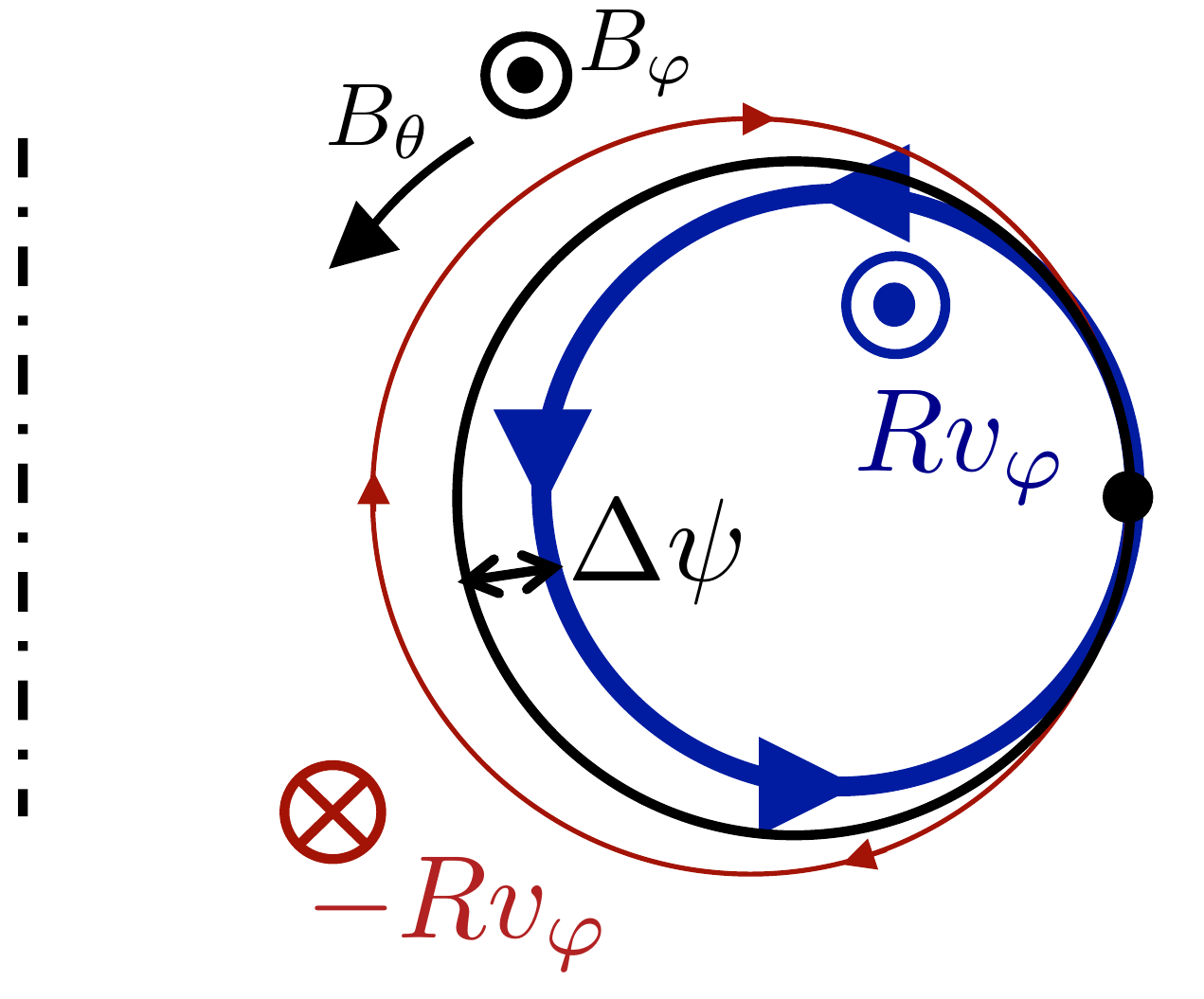}
\caption{Sketch of the origin of the diamagnetic particle flow. The poloidal cross section of a flux surface is represented by the black solid circle. The particles with positive parallel velocity drift radially inward (blue circle), while the particles with the negative parallel velocity drift radially outward (red circle). Because the plasma at inner radii is denser and hotter than the plasma at outer radii, the total parallel flux at the outer mid plane is positive and the ion diamagnetic flow is $V_{\|,d} \sim \Delta \psi ({\partial \ln p_i}/{\partial \psi}) v_{ti} \sim \rho_\star ({B}/{B_\theta}) v_{ti}$.  
}
\label{fig:rotation_dia}
\end{figure}

\subsection{Higher order gyrokinetic equations}\label{sec:high_GKE}

The symmetry in (\ref{f_tb1_sym}) and (\ref{phi_tb1_sym}) is broken by the effect of higher order terms in $\rho_{\star,\theta}$. One thus needs the higher order gyrokinetic equations to evaluate the momentum transport (i.e. $\Pi_1=0$ and $\Pi_2\neq0$).

In the gyrokinetic equations in this section, we only keep the corrections related to the diamagnetic flow and neglect all other corrections in $\rho_{\star,\theta}$ for simplicity. As shown in \cite{Parra:2014}, they are the dominant terms when the perpendicular characteristic length of the turbulence $l_\perp$ is small compared to the characteristic poloidal gyroradius $ (B/B_\theta)\rho_i $. The higher order ion gyrokinetic equation with both the diamagnetic flow and the E$\times$B flow in the lab frame is 
\begin{eqnarray}
&\frac{\partial f^{tb}_i}{\partial t}- \frac{c}{B} \left[\frac{\partial \phi^{bg}_0 }{\partial \psi}(\psi_0)+(\psi-\psi_0)\frac{\partial^2 \phi^{bg}_0 }{\partial \psi^2}(\psi_0)\right]\left(\nabla\psi \times \mathbf{\hat{b}}\right) \cdot  \nabla f^{tb}_{i}
\nonumber\\ &+\left(v_\| \mathbf{\hat{b}}+\mathbf{v}_M- \frac{c}{B} \nabla(\phi^{bg}_1+\langle\phi^{tb} \rangle) \times \mathbf{\hat{b}}\right) \cdot  \nabla f^{tb}_i\nonumber\\ &-\frac{Z_ie}{m_i}  (v_\| \mathbf{\hat{b}}+\mathbf{v}_M) \cdot\nabla \phi^{bg}  \frac{\partial f^{tb}_i}{\partial \mathcal{E}}  =\frac{c}{B} \nabla\langle\phi^{tb} \rangle \times \mathbf{\hat{b}} \cdot  \nabla f^{bg}_i\nonumber\\ &+ \frac{Z_ie}{m_i}( v_\| \mathbf{\hat{b}}+\mathbf{v}_M)\cdot\nabla\langle\phi^{tb} \rangle\frac{\partial f^{bg}_i}{\partial \mathcal{E}}+ \langle C(f_i) \rangle \label{GKE2},
\end{eqnarray}
where the distribution functions and potentials include the higher order corrections; i.e. $f^{tb}_i=f^{tb}_{1,i}+f^{tb}_{2,i}$, $f_i^{bg}=f_{0,i}^{bg}+f^{bg,E}_{1,i}+f^{bg,d}_{1,i}$, $\phi^{bg}=\phi_0^{bg}+\phi_1^{bg}$ and $\phi^{tb}=\phi^{tb}_{1}+\phi^{tb}_{2}$. Note that in the second term on the left hand side of the equation, we have Taylor expanded ${\partial \phi^{bg}_0 }/{\partial \psi}$ to keep the dependence on $\psi$. Here, $f^{bg}_{0,i}=f_{M,i}$ is the lowest order background piece that is a Maxwellian distribution function as in (\ref{GKE1}), and $f^{bg,E}_{1,i}$ and $f^{bg,d}_{1,i}$ are the small deviations from the Maxwellian due to the E$\times$B flow and the diamagnetic flow, respectively. For intrinsic rotation, the E$\times$B flow is of the same order as the diamagnetic flow (i.e. $f^{bg,E}_{1,i}\sim f^{bg,d}_{1,i}\sim  \rho_{\star,\theta}  f^{bg}_{0,i}$). For the E$\times$B flow, the correction to the background distribution function simply results in a shift of the background Maxwellian by the E$\times$B flow (i.e. $f_{0,i}^{bg}+f^{bg,E}_{1,i}= f_{M,i}(\mathbf{v}-\Omega_{\varphi,E}R{\hat{\boldsymbol{\varphi}}})$), making  $f^{bg,E}_{1,i} \simeq ({m_iv_\|}/{T_i})({I\Omega_{\varphi,E}}/{B})f_{M,i}$ where $\Omega_{\varphi,E}=-c({\partial  \phi_0^{bg}(\psi)}/{\partial \psi})$. The small deviation due to the diamagnetic flow, $f^{bg,d}_{1,i}$, is the neoclassical distribution function \cite{hinton1976theory,Helander:Coll}, which will be discussed in the next section. The higher order neoclassical potential $\phi_1^{bg}(\psi, \theta)$, which is determined by (\ref{DKE2}), is also considered in the above gyrokinetic equation (\ref{GKE2}). 

The quasineutrality condition in (\ref{QN}) must be corrected to include the higher order pieces $f^{tb}_{2,i}$ and $\phi^{tb}_{2}$, but the neoclassical corrections to quasineutrality are negligible for $B_\theta/B \ll \rho_i/l_\perp  \ll 1$ (see \cite{Parra:2014} for the details). We consider this limit, thus can use (\ref{QN}). 

 From the size of the correction to the background piece, we obtain the size of the corresponding corrections to the fluctuating pieces, $f^{tb}_{2,i}\sim  \rho_{\star,\theta} f^{tb}_{1,i}$ and $ \phi^{tb}_2\sim  \rho_{\star,\theta}  \phi^{tb}_{1}$. For the transformations $v_\| \rightarrow -v_\|$, $\theta \rightarrow -\theta$ and $k_\psi \rightarrow -k_\psi$, all additional higher order terms in the gyrokinetic equation have the parity opposite to the lowest order terms in (\ref{GKE1}). The parity of the higher order corrections to the distribution function and potential is obtained by linearizing the effect of the corrections small in $ \rho_{\star,\theta} $ (compare the following with (\ref{f_tb1_sym}) and (\ref{phi_tb1_sym})),  
\begin{eqnarray}
\hat{f^{tb}_{2,i}}(\theta,v_\|,k_\psi)= \hat{f^{tb}_{2,i}}(-\theta,-v_\|,-k_\psi)\label{f_tb2_sym} \\ \hat{\phi}_2^{tb}(\theta,k_\psi)= \hat{\phi}_2^{tb} (-\theta,-k_\psi)\label{phi_tb2_sym}.
\end{eqnarray}
The different symmetry of $\hat{f^{tb}_{2,i}}$ and $\hat{\phi}_2^{tb}$ results in non-vanishing higher order momentum transport,
\begin{eqnarray} 
\pi_{\|2}(\theta,v_\|,k_\psi)= \pi_{\|2} (-\theta,-v_\|,-k_\psi),\label{pi_even2}
\end{eqnarray}
where $ \pi_{\|2}\propto Re[ i k_\alpha (\hat{f^{tb}_{2,i}} \hat{\phi_1^{tb}} ^*+\hat{f^{tb}_{1,i}} \hat{\phi_2^{tb}} ^*)v_\|]$.


\subsection{Neoclassical correction to Maxwellian equilibria}
 

The correction $f_{1,s}^{bg,d}$ to the Maxwellian distribution function in (\ref{GKE2}) is derived from the drift kinetic equation for species $s$ \cite{hinton1976theory,Helander:Coll},
\begin{eqnarray}
v_\| \hat{\mathbf{b}}\cdot \nabla H_{1,s} + \mathbf{v}_{M,s}\cdot \nabla f_{0,s}^{bg}=C_s(H_{1,s}), \label{DKE0}
\end{eqnarray}
where  $H_{1,s}=f^{bg,d}_{1,s}+({Z_se\phi_1^{bg}}/{T_s}) f_{0,s}^{bg}$. Here, the lowest order background potential $\phi_0^{bg}$ which results in the E$\times$B flow is not included in (\ref{DKE0}), but the higher order background potential $\phi_1^{bg}$ due to the neoclassical correction is included. The lowest order potential $\phi_0^{bg}$ is considered separately in $f^{bg,E}_{1,s}$ of (\ref{GKE2}). Using $\mathbf{v}_{M,s}\cdot \nabla f_{0,s}^{bg}=I v_\| \hat{\mathbf{b}}\cdot \nabla ({v_\|}/{\Omega_s}) ({\partial f_{0,s}^{bg}}/{\partial \psi})$, the drift kinetic equation is simplified to 
\begin{eqnarray}
v_\| \hat{\mathbf{b}}\cdot\nabla \left(H_{1,s}-f^M_{1,s}\right)=C_s(H_{1s}),
\label{DKE1}
\end{eqnarray}
where $f^M_{1,s}=-({Iv_\|}/{\Omega_s})\left[{\partial \ln n_s}/{\partial \psi}+({m_sv^2}/{2T_s}-{3}/{2})({\partial \ln T_s}/{\partial \psi})\right]  f_{0,s}^{bg}$. Depending on collisionality, different solutions for $H_{1,s}$ are obtained. By employing the quasineutrality condition, 
\begin{eqnarray}
\sum_s Z_s \int d^3v \left(H_{1,s}-\frac{Z_se\phi_1^{bg}}{T_s} f_{0,s}^{bg}\right)=0, \label{DKE2}
\end{eqnarray}
we can also obtain $\phi_1^{bg}$. 

To study the symmetry breaking mechanisms due to the higher order background distribution function, the solution of the ion drift kinetic equation is decomposed into
\begin{eqnarray}
f^{bg,d}_{1,i}=f^{bg,V_\|}_{1,i}+f^{bg,q_\|}_{1,i}+f^{bg,other}_{1,i}\sim  \rho_{\star,\theta}  f_{0,i}^{bg},\label{neo1}
\end{eqnarray}
where $f^{bg,V_\|}_{1,i}$ is the piece resulting in the diamagnetic parallel particle flow, $f^{bg,q_\|}_{1,i}$ is the piece giving the diamagnetic parallel heat flow, and $f^{bg,other}_{1,i}$ accounts for all other contributions. 

The piece $f^{bg,V_\|}_{1,i}$ is
\begin{eqnarray}
f^{bg,V_\|}_{1,i}=\frac{m_iv_\|V_{\|,d}}{T_i}f^{bg}_0\label{neo2}
\end{eqnarray}
where $V_{\|,d}\left({\partial p_i}/{\partial \psi},{\partial T_i}/{\partial \psi}, \nu_\star.. .\right)$ is the diamagnetic parallel particle flow that depends on the pressure gradient, temperature gradient and collisionality $\nu_\star$. Here, $\nu_\star \equiv ({\nu_{ii}qR}/{\epsilon^{3/2}v_{ti}})$ is the ratio of effective collision frequency for pitch angle scattering to the bounce frequency, $\nu_{ii}$ is the ion-ion collision frequency, and $\epsilon=r/R$ is the inverse of the aspect ratio. To lowest order, for a plasma with only one ion species, the diamagnetic parallel particle flow is given by
\begin{eqnarray}
V_{\|,d}=\frac{I(\psi)}{B}\Omega_{\varphi,p}(\psi)+\frac{K(\psi)}{n_i(\psi)}B,\label{general_Vpar}
\end{eqnarray}
where $\Omega_{\varphi,p}=-({c}/{Z_ien_i})({\partial p_i}/{\partial \psi})$ and $K(\psi)\propto {\partial  T_i}/{\partial \psi}$. For example, the diamagnetic parallel particle flow in the banana regime $\nu_\star \ll 1$ for concentric circular flux surfaces and large aspect ratio is
\begin{eqnarray}
V_{\|,d}=-\frac{IT_i}{m_i\Omega_{i0}}\left[\frac{1}{h}\frac{\partial \ln p_i}{\partial \psi}-1.17 {f_c}h\frac{\partial \ln T_i}{\partial \psi}\right],\label{Vpar_Banana}
\end{eqnarray}
where $h={B}/{B_0}$, $B_0$ and $\Omega_{i0}$ are the magnetic field magnitude and ion gyrofrequency at the magnetic axis, respectively, and $f_c$ is the effective fraction of passing particles \cite{Helander:Coll}. 

Combining the parallel diamagnetic flow with the parallel flow due to the lowest order radial electric field and the perpendicular flow, the total equilibrium flow in a tokamak is obtained \cite{Rosenbluth:NF1996},
\begin{eqnarray}
\mathbf{V_i}&=&(V_{\|,d}+V_{\|,E})\mathbf{\hat{b}}+\mathbf{V_\perp}\nonumber\\
&=&(\Omega_{\varphi,p}(\psi)+\Omega_{\varphi,E}(\psi)) R \hat{\boldsymbol{\varphi}} + \frac{K(\psi)}{n_i(\psi)} \mathbf{B}, \label{neo_flow}
\end{eqnarray}
where $V_{\|,E}=-({cI}/{B})({\partial \phi_0^{bg}}/{\partial \psi})$ is the parallel flow due to the lowest order radial electric field. Here, $\mathbf{V_\perp}=-(\Omega_{\varphi,p}+\Omega_{\varphi,E})({\mathbf{B}\times \nabla \psi}/{B^2})$ is the perpendicular flow due to the E$\times$B and the pressure driven diamagnetic drift, and we use $\mathbf{\hat{b}}\times \nabla \psi=I\mathbf{\hat{b}}-BR\hat{\boldsymbol{\varphi}}$. Notice that the poloidal flow is determined only by the term proportional to $K(\psi)$, while the toroidal flow is given by $\Omega_{\varphi,p} +\Omega_{\varphi,E} $ and the term proportional to $K(\psi)$. Thus the toroidal particle flow is
\begin{eqnarray}
V_\varphi&=&\mathbf{V_i} \cdot \mathbf{\hat{\boldsymbol{\varphi}}}\equiv \Omega_{\varphi} R\nonumber\\
&=&(\Omega_{\varphi,p} (\psi) +\Omega_{\varphi,E} (\psi) +\Omega_{\varphi, T} (\psi, \theta) ) R\label{neo_flow}
\end{eqnarray}
where $\Omega_{\varphi, T} (\psi, \theta) = {K(\psi)I(\psi)}/({n_i(\psi) R^2(\psi, \theta)})$. We can divide the toroidal angular frequency of the temperature gradient driven flow into two pieces, $\Omega_{\varphi, T} (\psi, \theta) =\overline{\Omega_{\varphi, T}} (\psi)+\Delta \Omega_{\varphi, T} (\psi, \theta)$. The piece contributing to the  angular momentum is
\begin{eqnarray}
\overline{\Omega_{\varphi, T}} (\psi)&=& \frac{\langle\Omega_{\varphi, T} R^2 \rangle_\psi (\psi)}{\langle R^2 \rangle_\psi (\psi)}=\frac{K(\psi)I(\psi)}{n_i(\psi)}\frac{\int_0^{2\pi} d\theta (\mathbf{B} \cdot \nabla \theta)^{-1}}{\int_0^{2\pi} d\theta R^2(\mathbf{B} \cdot \nabla \theta)^{-1}},
\end{eqnarray}
and the piece not contributing to the toroidal angular momentum is
 \begin{eqnarray}
\Delta \Omega_{\varphi, T} (\psi, \theta) &=&\frac{K(\psi)I(\psi)}{n_i(\psi)} \left(\frac{1}{R^2(\psi, \theta)}-\frac{\int_0^{2\pi} d\theta (\mathbf{B} \cdot \nabla \theta)^{-1}}{\int_0^{2\pi} d\theta R^2(\mathbf{B} \cdot \nabla \theta)^{-1}}\right),\label{defdelOmegaT}
\end{eqnarray}
where $\langle  \Delta \Omega_{\varphi, T} R^2 \rangle_\psi=0$.
Then, the particle flow is given by 
\begin{eqnarray}
V_\varphi&=&(\Omega_{\varphi,d} (\psi) +\Omega_{\varphi,E} (\psi) )R +\Delta \Omega_{\varphi, T} (\psi, \theta) R \label{neo_flow}
\end{eqnarray}
where $\Omega_{\varphi,d} (\psi)=\Omega_{\varphi,p} (\psi) + \overline{\Omega_{\varphi, T}} (\psi)$.

The diamagnetic parallel heat flow is due to the terms proportional to $v_\|v^2/2$ in the distribution function, $f^{bg,d}_{1,i}$. The parallel heat flow breaks the symmetry even in the absence of rotation because it corresponds to a distribution function that is odd in $v_\|$,
\begin{eqnarray}
f^{bg,q_\|}_{1,i}=\frac{2}{5}\frac{m_iv_\|q_\|}{p_iT_i}\left(\frac{m_iv^2}{2T}-\frac{5}{2}\right)f^{bg}_0,\label{neo_qpar}
\end{eqnarray}
where $p_i$ is the ion pressure and the heat flow $q_\|\left({\partial T_i}/{\partial \psi}, \nu_\star.. .\right)$ is also a function of the temperature gradient and collisionality. We can see that (\ref{neo2}) and  (\ref{neo_qpar}) satisfy $\int dv_\| v_\| f^{bg,V_\|}_{1,i}=V_{\|,d}$, $\int dv_\| v_\| f^{bg,q_\|}_{1,i}=0$, $\int dv_\| v_\|(m_iv^2/2T_i-5/2) f^{bg,V_\|}_{1,i}=0$ and $\int dv_\| v_\|(m_iv^2/2T_i-5/2) f^{bg,q_\|}_{1,i}=q_\|$. 

 \subsection{Symmetry breaking due to diamagnetic effects}

The two types of toroidal rotation ($\Omega_{\varphi,d}$ and $\Omega_{\varphi,E}$) have different origins and characteristics. The rotation $\Omega_{\varphi,E}$ is proportional to the radial electric field because of radial force balance, and as a result the radial electric field changes in the momentum confinement time scale \cite{Parra:NF2011}. The electric field changes the energy and orbits of each particle. Conversely, the pressure and the temperature gradients in $\Omega_{\varphi,d}$ are determined by the turbulent anomalous transport of particles and energy, changing on the energy confinement time scale. The pressure and the temperature cannot change the energy and orbits of individual particles. These different characteristics are reflected in the gyrokinetic equation (\ref{GKE2}) by the different terms for $\Omega_{\varphi,d}$ and $\Omega_{\varphi,E}$. There are additional terms on the left hand side of (\ref{GKE2}) related to $\phi_0^{bg}$ (compare to (\ref{GKE1})). The term $- ({c}/{B})(\psi-\psi_0)({\partial^2 \phi^{bg}_0 }/{\partial \psi^2})(\nabla\psi\times \mathbf{\hat{b}}) \cdot  \nabla f_i^{tb}$ is the effect of the background E$\times$B drift. The other term, $-({Z_ie}/{m_i})(\mathbf{v}_M \cdot\nabla \phi_0^{bg}) ({\partial f^{tb}_i}/{\partial \mathcal{E}})$, is the acceleration of the particles in the radial electric field when the particles change their radial positions due to the curvature and $\nabla B$ drift. The diamagnetic flow does not have terms like these on the left hand side because it cannot change the orbits and energy of the particles.

Due to their different characteristics, the effects of the two flows $\Omega_{\varphi,E}$ and $\Omega_{\varphi,d}$ on the turbulent momentum transport are different. Then, in the low flow regime, it is convenient to write the ion toroidal momentum transport as  
\begin{eqnarray}
 &&\Pi=\Pi_{{int}}^\prime- n_im_i \langle R^2\rangle_\psi\left(P_{\varphi,d}  \Omega_{\varphi,d}+P_{\varphi,E} \Omega_{\varphi,E} \right)- n_im_i \langle R^2 P_{\varphi,\theta}  \Delta \Omega_{\varphi, T}\rangle_\psi\nonumber \\&&-  n_im_i \langle R^2\rangle_\psi \left(\chi_{\varphi,d} \frac{\partial\Omega_{\varphi,d}}{\partial r} +\chi_{\varphi,E}\frac{\partial\Omega_{\varphi,E}}{\partial r} \right)-  n_im_i\left \langle R^2\chi_{\varphi,\theta}\frac{\partial\Delta\Omega_{\varphi,T}}{\partial r}\right \rangle_\psi \label{Pi20},
 \end{eqnarray}
where $\Pi_{{int}}^\prime$ is the redefined intrinsic momentum flux for $\Omega_{\varphi,d}=0$, $\Omega_{\varphi,E}=0$, $\Delta \Omega_{\varphi,T}=0$, ${\partial \Omega_{\varphi,d}}/{\partial r}=0$, ${\partial \Omega_{\varphi,E}}/{\partial r}=0$, and ${\partial \Delta\Omega_{\varphi,T}}/{\partial r}=0$. Equation (\ref{Pi20}) is valid for $\Omega_{\varphi,d}$, $\Omega_{\varphi,E}$, and $\Delta\Omega_{\varphi,T}$ sufficiently small to be able to linearize $\Pi$ around  $\Omega_{\varphi,d}=\Omega_{\varphi,E}=\Delta\Omega_{\varphi,T}=0$. 
By using $\langle R^2 \Omega_\varphi\rangle_\psi = \langle R^2\rangle_\psi(\Omega_{\varphi,d}+\Omega_{\varphi,E})$, the original model in (\ref{Pi_com1}) is recovered by
\begin{eqnarray}
\Pi&=&\Pi_{{int}}
 -n_im_i \langle R^2 \rangle_\psi P_{\varphi,E} \left(\Omega_{\varphi,d}+\Omega_{\varphi,E}\right)\nonumber \\&&-n_i m_i\langle R^2\rangle_\psi \chi_{\varphi,E}\left(\frac{\partial\Omega_{\varphi,d}}{\partial r}+\frac{\partial\Omega_{\varphi,E}}{\partial r}\right).\label{Pi2}
\end{eqnarray}
Then, equations (\ref{Pi20}) and (\ref{Pi2}) result in the relation between $\Pi_{{int}}$ and  $\Pi_{{int}}^\prime$  
\begin{eqnarray}
\Pi_{{int}}&&= \Pi_{{int}}^\prime-n_im_i \langle R^2 \rangle_\psi \Delta P_{\varphi} \Omega_{\varphi,d}- n_im_i \langle R^2 P_{\varphi,\theta}  \Delta \Omega_{\varphi, T}\rangle_\psi \nonumber \\&&-n_im_i\langle R^2\rangle_\psi\Delta \chi_{\varphi}\frac{\partial\Omega_{\varphi,d}}{\partial r} -  n_im_i\left \langle R^2\chi_{\varphi,\theta}\frac{\partial\Delta\Omega_{\varphi,T}}{\partial r}\right \rangle_\psi,\label{Pi3}
\end{eqnarray}
where $\Delta P_{\varphi}= P_{\varphi,d}-P_{\varphi,E}$ and $\Delta \chi_{\varphi}=\chi_{\varphi,d}-\chi_{\varphi,E}$. Here, $\Pi_{{int}}^\prime$ includes the effect of the parallel heat flux, $f_{1,i}^{bg,q_\|}$, of $f_{1,i}^{bg,other}$ in (\ref{neo1}), and of the neoclassical potential $\phi_1^{bg}$. 

 The intrinsic momentum transport in (\ref{Pi3}) is then
\begin{eqnarray}
\Pi_{{int}}= \Pi_{{int}}^{\Delta  P_{\varphi}}+ \Pi_{{int}}^{\Delta  \chi_{\varphi}}+\Pi_{{int}}^{\theta}+\Pi_{{int}}^{q_\|}+\Pi_{{int}}^{\phi_1^{bg}}+\Pi_{{int}}^{other},\label{Pi4}
\end{eqnarray}
where the contribution of the difference in the momentum pinches is 
\begin{eqnarray}
 \Pi_{{int}}^{\Delta  P_{\varphi}}=-n_im_i \langle R^2\rangle_\psi \Delta P_{\varphi} \Omega_{\varphi,d}, \label{PidelP}
 \end{eqnarray}
the contribution of the different diffusivities is
 \begin{eqnarray}
 \Pi_{{int}}^{\Delta  \chi_{\varphi}}= -n_i m_i\langle R^2\rangle_\psi\Delta \chi_{\varphi}\frac{\partial\Omega_{\varphi,d}}{\partial r}, \label{Pidelchi}
 \end{eqnarray}
 and the contribution of the different poloidal dependence of the temperature driven diamagnetic flow is
   \begin{eqnarray}
\Pi_{{int}}^{\theta}=-n_im_i \langle R^2 P_{\varphi,\theta}  \Delta \Omega_{\varphi, T}\rangle_\psi  -n_im_i\left \langle R^2\chi_{\varphi,\theta}\frac{\partial\Delta\Omega_{\varphi,T}}{\partial r}\right \rangle_\psi. \label{Pitheta}
 \end{eqnarray}
Here, $\Pi_{{int}}^{q_\|}$ and $\Pi_{{int}}^{\phi_1^{bg}}$ are the intrinsic momentum transport due to the symmetry breaking caused by the parallel heat flow and the neoclassical potential, satisfying $f^{bg,q_\|}_{1,i}(\theta, v_\|)=-f^{bg,q_\|}_{1,i}(-\theta,-v_\|)$ and $\phi_1^{bg}(\theta)=-\phi_1^{bg}(-\theta)$, respectively. 
Finally, $\Pi_{{int}}^{other}$ includes the contribution due to the piece $f^{bg,other}_{1,i}$ breaking the symmetry of the turbulence, because it satisfies  $f^{bg,other}_{1,i}(\theta,v_\|)=-f^{bg,other}_{1,i}(-\theta,-v_\|)$. 

Thus, the intrinsic momentum transport for zero flow and zero flow shear is a function of $\Omega_{\varphi,d}$, ${\partial\Omega_{\varphi,d}}/{\partial r}$,  $\Delta \Omega_{\varphi, T}$, $q_\|$ and $\phi_1^{bg}$. In general, the diamagnetic flow depends on the pressure gradient, the temperature gradient, and collisionality, so the intrinsic momentum transport is
\begin{eqnarray}
\Pi_{{int}}= \Pi_{{int}}\left(\frac{\partial p_i}{\partial \psi},\frac{\partial T_i}{\partial \psi}, \nu_\star,\frac{\partial^2 p_i}{\partial \psi^2},\frac{\partial^2 T_i}{\partial \psi^2}, \frac{\partial \nu_\star}{\partial \psi},\Pi_{{int}}^\prime\right),\label{Pi_int0}
\end{eqnarray}
In the next section, we will show the contribution of each term to the momentum flux by evaluating  $\Pi_{{int}}^{\Delta  P_{\varphi}}$, $\Pi_{{int}}^{\Delta  \chi_{\varphi}}$, $\Pi_{{int}}^{\theta}$, $\Pi_{{int}}^{q_\|}$, and $\Pi_{{int}}^{\phi_1^{bg}}$ with gyrokinetic simulations.

\section{Evaluation of intrinsic momentum transport}\label{sec:int_mom}

We used the gyrokinetic code GS2 \cite{Dorland:PRL2000} to evaluate the intrinsic momentum flux. For numerical reasons, the gyrokinetic equation is solved in the frame rotating with $\Omega_{\varphi,E}$, thereby avoiding the energy derivative terms due to $\phi_0^{bg}$ on the left hand side of (\ref{GKE2}). The kinetic equation in the frame rotating with $\Omega_{\varphi,E}(\psi_0)$
\begin{eqnarray}
&\frac{\partial f_i^{tb(R)}}{\partial t^\prime}+(\psi-\psi_0)\frac{\partial \Omega_{\varphi,E}}{\partial \psi}R \hat{\boldsymbol{\varphi}}\cdot \nabla  f_i^{tb(R)}\nonumber\\&+\left(v_\|^\prime \mathbf{\hat{b}}+ \mathbf{v}_M^\prime+\mathbf{v}_{C,\Omega_{\varphi,E}}^\prime-\frac{c}{B} \nabla(\phi_1^{bg}+\langle\phi^{tb} \rangle)\times \mathbf{\hat{b}} \right) \cdot \nabla f_i^{tb(R)}-\frac{Z_ie}{m_i}  v_\|^{\prime}(  \nabla \phi_1^{bg} \cdot \mathbf{\hat{b}})\frac{\partial f^{tb(R)}_i}{\partial \mathcal{E}^{\prime}}\nonumber\\ &=\frac{c}{B} \nabla\langle\phi^{tb} \rangle \times \mathbf{\hat{b}} \cdot  \nabla f_{i}^{bg(R)}+\frac{c}{B} \nabla\langle\phi^{tb} \rangle \times \mathbf{\hat{b}} \cdot  \nabla \psi \frac{m_iIv_\|^\prime}{T_iB}\frac{\partial \Omega_{\varphi,E}}{\partial \psi }f_{i}^{bg(R)}\nonumber\\&+ \frac{Z_ie}{m_i}(  v_\|^\prime\mathbf{\hat{b}}+\mathbf{v}_M^\prime+\mathbf{v}^\prime_{C,\Omega_{\varphi,E}})  \cdot\nabla\langle\phi^{tb} \rangle\frac{\partial f_{i}^{bg(R)}}{\partial \mathcal{E}^\prime}+\langle C(f_i)\rangle\label{GKE4},
\end{eqnarray}
where ${\partial }/{\partial t^\prime}={\partial }/{\partial t}+ \Omega_{\varphi,E}(\psi_0)R \hat{\boldsymbol{\varphi}}\cdot \nabla$ is the Doppler shifted time derivative, $\mathbf{v}_{C,\Omega_{\varphi,E}}^\prime=({2v_\|^\prime\Omega_{\varphi,E}}/{\Omega_i})\mathbf{\hat{b}} \times [(\nabla{R} \times \hat{\boldsymbol{\varphi}})\times \mathbf{\hat{b}}]$ is the Coriolis drift and $f_{i}^{bg(R)}=f_{0,i}^{bg}(v_\|^\prime)+f_{1,i}^{bg,d}(v_\|^\prime)$
is the perturbed Maxwellian distribution function due to the diamagnetic effects in the rotating frame. 
 Here, we apply a transformation from the lab frame velocity $\mathbf{v}$ to the rotating frame velocity $\mathbf{v}^{\prime}$ using the relation $\mathbf{v}^{\prime}=\mathbf{v}- R\Omega_{\varphi,E} \hat{\boldsymbol{\varphi}}$. The parallel velocity becomes $ v_\|^{\prime}= v_\|- I\Omega_{\varphi,E}/B$ and the kinetic energy becomes $\mathcal{E}^\prime=\mathcal{E}-I\Omega_{\varphi,E} v_\|/B$ \cite{Parra:NF2011}. The superscript $(R)$ indicates the value in the rotating frame, whereas no superscript  denotes the value in the lab frame. The conversion of the gyrokinetic equation between the laboratory frame and the rotating frame is given in appendix A of \cite{Parra:NF2011} for $\rho_\star \ll B_\theta/B \ll 1$.
 

To investigate the intrinsic momentum transport in a non-rotating state, we set the toroidal flow to be zero,  $\Omega_{\varphi}=\Omega_{\varphi,d} +\Omega_{\varphi,E}=0$. The intrinsic momentum transport is the flux in the lab frame, which is 
\begin{eqnarray}
\Pi_{{int}}=\Pi^{tb}=\Pi^{tb(R)}+m_i\Omega_{\varphi,E}\langle \Gamma R^2 \rangle_\psi=\Pi^{tb(R)}-m_i\Omega_{\varphi,d}\langle \Gamma R^2 \rangle_\psi , \label{Pi0}
\end{eqnarray}
where $\Gamma$ is the particle flux.

For the gyrokinetic simulations in this paper, we used two species (deuterons and electrons), and the plasma parameters of the Cyclone Base Case that model experimental conditions in the plasma core : $R_0/L_{T}=6.9$, $R_0/L_{n}=2.2$, $q=1.4$, $r/a=0.54$, $R_0/a=3.0$ and $\hat{s}=0.8$. Here, $L_{T}=-T_i/(dT_i/dr)=-T_e/(dT_e/dr)$ and $L_{n}=-n/(dn/dr)$ are the characteristic lengths of temperature and density, respectively, $R_0\simeq\langle R \rangle_\psi$, $q$ is the safety factor, and $\hat{s}$ is the magnetic shear.  Due to the new higher order terms in (\ref{GKE4}), we need to specify values for $\rho_\star$, $d(R_0/L_n)/dr$ and $d(R_0/L_T)/dr$.  We choose $\rho_\star=0.01$\footnotemark[1],  $d(R_0/L_n)/dr=0$ and $d(R_0/L_T)/dr=0$.

\footnotetext[1]{In many experiments, $0.001<\rho_\star<0.01$}

The GS2 simulations use 32 grid points in the parallel coordinate $\theta$, 12 grid points in kinetic energy, and 20 grid points in pitch angle. The box size of the simulation in both radial $(\psi)$ and binormal ($\alpha$) directions is approximately 125 $\rho_i$.  We use 128 and 22 Fourier modes in the radial and binormal directions, respectively.

Neoclassical effects depend strongly on collisionality $\nu_\star$. We investigate the low collisionality intrinsic momentum transport due to neoclassical particle and heat flows in section 4.1. We compare these results to results for moderate collisionality in section 4.2. In the moderate collisionality case, the contributions of $f^{bg,other}_{1,i}$ and $\phi_{1}^{bg}$ to the momentum transport are non-negligible, whereas they are not important for low collisionality.

\clearpage
\subsection{Low collisionality}
Table 1 shows GS2 results for the contributions of each neoclassical correction to the momentum flux for low collisionality $\nu_\star \ll 1$. For all the contributions, the normalized radial flux of toroidal angular momentum divided by the normalized turbulent heat flux is of order the poloidal rhostar $(B/B_\theta)\rho_\star\sim 0.1$, as expected. The neoclassical parallel particle and heat flows and their radial gradients in figure 2 are used to obtain the simulation results in Table 1. In the low collisionality case, the total momentum flux is almost the summation of the contributions from the particles and heat flows, which are comparable with opposite signs (i.e. $\Pi_{{int}} \simeq\Pi_{{int}}^{V_\|} + \Pi_{{int}}^{q_\|}$ from Table 1-(a), (b) and (c)). Other contributions to the momentum flux in Table 1-(d) and (e) are clearly smaller. 

\begin{figure} 
(a) \\
 \includegraphics[scale=0.5]{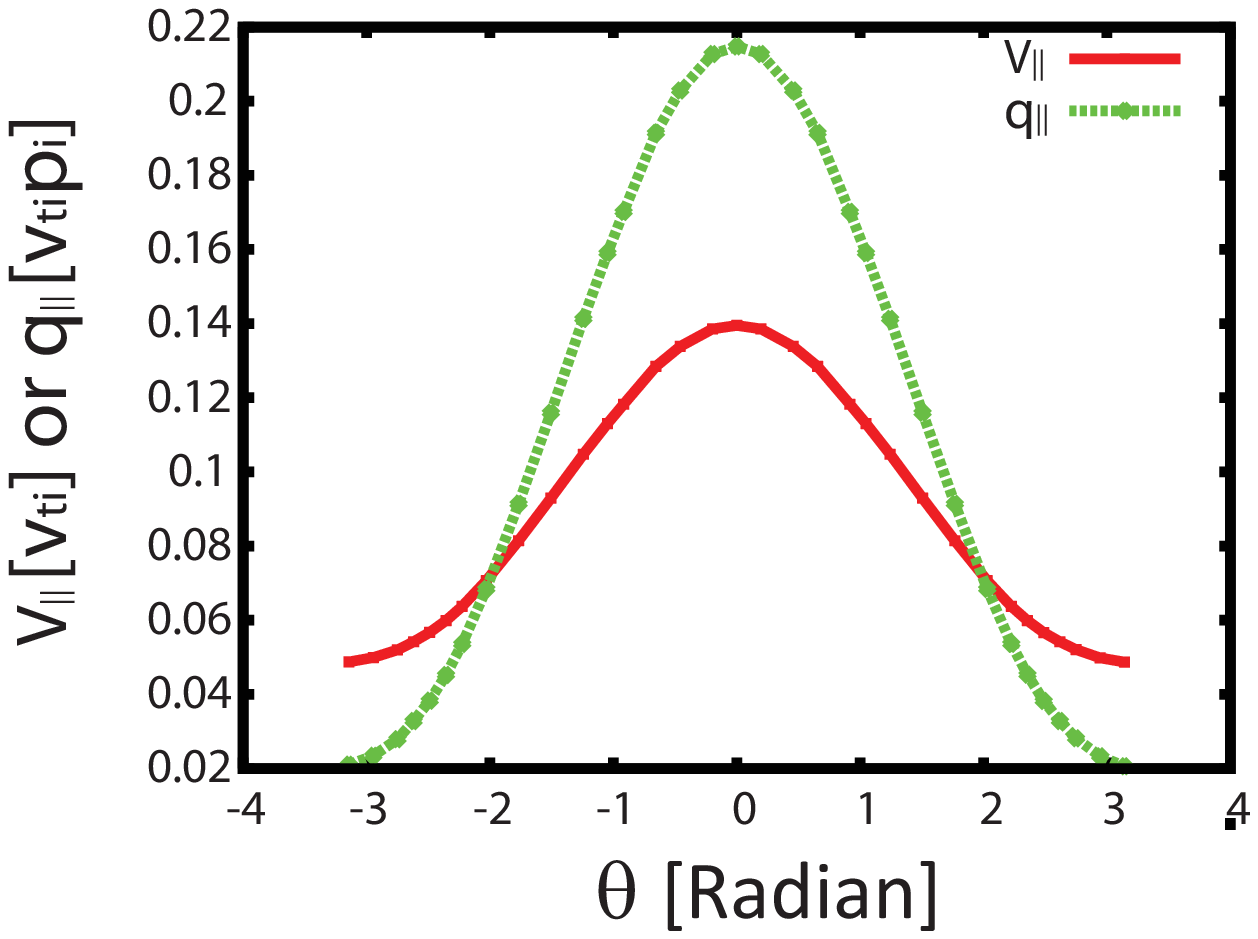}\\
(b)\\
 \includegraphics[scale=0.5]{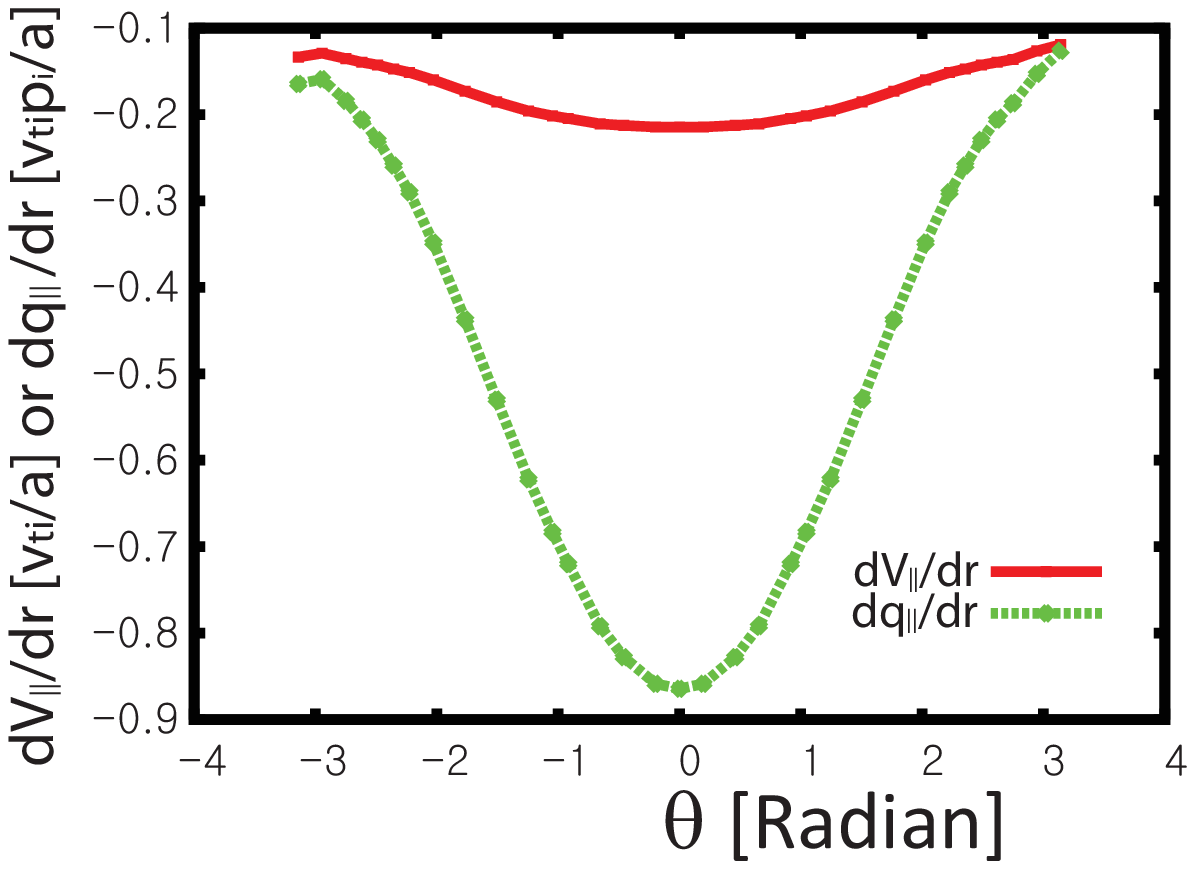}
\caption{(a) The normalized diamagnetic parallel particle flow (red) $V_{\|,d}/v_{ti}$ and the parallel heat flow $q_\|/v_{ti}p_i$ (green) in terms of poloidal angle. (b) The normalized radial gradient of diamagnetic parallel particle flow (red) $(\partial V_{\|,d}/\partial r)/(v_{ti}/a)$ and the parallel heat flow $(\partial q_\|/\partial r)/(v_{ti}p_i/a)$ (green) in terms of poloidal angle. All profiles are obtained with the neoclassical code NEO \cite{belli2008kinetic,belli:PPCF2012} for the Cyclone Base Case with $\rho_\star =0.01$, $d(R_0/L_n)/dr=0$, $d(R_0/L_T)/dr=0$ and low collisionality $\nu_\star =0.043$.}
\label{fig:rotation_dia}
\end{figure}

\Table{\label{tab:pinch_diff}The intrinsic radial flux of toroidal angular momentum $\Pi_{int}$ (we assume a non-rotating state $\Omega_{\varphi}=\Omega_{\varphi,d} +\Omega_{\varphi,E}=0$), divided by the ion heat flux $Q_i$ given by GS2 for the Cyclone Base Case with $\rho_\star =0.01$, $d(R_0/L_n)/dr=0$, $d(R_0/L_T)/dr=0$ and low collisionality $\nu_\star =0.043$ for (a) full neoclassical function and potential given by NEO, (b) only the particle flow piece $f^{bg,V_\|}_{1,i}$, (c) only the parallel heat flow piece $f_{1,i}^{bg,q_\|}$ given by NEO, (d) only the neoclassical potential and (e) only the remaining piece $f_{1,i}^{bg,other}$. Here, $\sigma\left(\frac{ \Pi_{int}}{Q_i}\frac{v_{ti}}{R_0}\right)$ is the standard deviation of the normalized momentum flux from the time averaged value due to turbulent fluctuations.}
\br
& \centre{1}{Size}& \centre{2}{GS2 results} \\
&  &$\frac{\Pi_{int}}{Q_i}\frac{v_{ti}}{R_0}$ &$\sigma\left(\frac{ \Pi_{int}}{Q_i}\frac{v_{ti}}{R_0}\right)$\\
\mr
(a) $\Pi_{{int}}$ due to $f_{1,i}^{bg,d}$ and $\phi_{1}^{bg}$ & $f_{1,i}^{bg,d}=f^{bg,V_\|}_{1,i}+f^{bg,q_\|}_{1,i}+f^{bg,other}_{1,i}$  & -0.022 & 0.018 \\
(b) $\Pi_{{int}}^{V_\|}$ due to $f_{1,i}^{bg,V_\|}$ &  $V_{\|,d}$ and $dV_{\|,d}/dr$ in figure 2 (red) &0.025  & 0.020\\
(c) $\Pi_{{int}}^{q_\|}$ due to $f_{1,i}^{bg,q_\|}$& $q_\|$ and  $dq_\|/dr$ in figure 2 (green)  &-0.046  & 0.015 \\
(d) $\Pi_{{int}}^{\phi_{1}^{bg}}$  due to $\phi_{1}^{bg}$ &$e \sqrt{\langle (\phi_{1}^{bg})^2 \rangle_\psi}/T_i \simeq$ 0.0005  & 0.002 & 0.019 \\
(e) $\Pi_{{int}}^{other}$ due to $f_{1,i}^{bg,other}$&$f_{1,i}^{bg,d}-f^{bg,V_\|}_{1,i}-f^{bg,q_\|}_{1,i}$ and $\phi_{1}^{bg}=0$& 0.001  & 0.016 \\
\br
\end{tabular}
\end{indented}
\end{table}

\subsubsection{Neoclassical parallel particle flow contribution.}\label{particle_flow}
The momentum transport due to the neoclassical parallel particle flow $\Pi_{{int}}^{V_\|}$ can be divided into three pieces: (1) flux due to the different pinches for the diamagnetic flow and the E$\times$B flow ($\Pi_{{int}}^{\Delta  P_{\varphi}}=-n_im_i \langle R^2\rangle_\psi \Delta P_{\varphi} \Omega_{\varphi,d}$), (2) flux due to the different momentum diffusivities for the diamagnetic flow and the E$\times$B flow ($\Pi_{{int}}^{\Delta  \chi_{\varphi}}= -n_i m_i\langle R^2\rangle_\psi\Delta \chi_{\varphi}({\partial\Omega_{\varphi,d}}/{\partial r})$), and (3) flux due to the different poloidal dependence of the pressure gradient driven flow and the temperature gradient driven flow ($\Pi_{{int}}^{\theta}$). The momentum transport due to the particle flow $\Pi_{{int}}^{V_\|}$ in Table 1-(b) is approximately the summation of the three pieces in Table 2-(a), (b) and (c). 

The pressure driven piece of the rotation is $\Omega_{\varphi,p}(\psi)R_0/v_{ti}=-({c}/{Z_ien_i})({\partial p_i}/{\partial \psi})(R_0/v_{ti})=(1/\sqrt{2})(q/(r/R_0))\rho_\star (a/R_0)^2(R_0/L_{n}+R_0/L_{T})=0.167$, as shown in the green curve of figure 3-(a). By solving the drift-kinetic equation in (\ref{DKE0}) for this case using the neoclassical code NEO \cite{belli2008kinetic,belli:PPCF2012}, we can obtain the piece of the particle flow that is proportional to the temperature gradient, $({K(\psi)}/{n_i(\psi)})B$ in (\ref{general_Vpar}), as shown in the blue curve of figure 3-(a). As shown in figure 3-(a), this piece tends to cancel the other piece of the particle flow $\Omega_{\varphi,p}(\psi)R_0$ in a low collisionality regime, giving ${\Omega_{\varphi,d} R_0}/{v_{ti}}=0.087$ and the strong poloidal variation $\Delta \Omega_{\varphi, T}$ in figure 3-(b). The red curves in figure 3-(a) and (b) correspond to the red curve in figure 2-(a), representing the particle flow $V_{\|,d}$ in which both contributions from $\Omega_{\varphi,d}$ and $\Delta \Omega_{\varphi, T}$ are included.  

The different pinch coefficients were investigated in \cite{Lee:NF2014} for the case with $R_0/L_{T}=9.0$, $R_0/L_{n}=9.0$, $q=2.5$. For these parameters, there is a $22\%$ difference between the pinch coefficients: ${P_{\varphi,d}}/{\chi_{\varphi,E}}\simeq 3.5/R_0$  and ${P_{\varphi,E}}/{\chi_{\varphi,E}}\simeq 2.9/R_0$. This difference in the pinch coefficients may explain the rotation peaking observed in H-mode \cite{Lee:NF2014}. For the Cyclone Base Case, the difference between the pinch coefficients is also non-negligible,
\begin{eqnarray}
 \frac{\Delta P_{\varphi}}{\chi_{\varphi,E}}\equiv \frac{P_{\varphi,d}-P_{\varphi,E}}{\chi_{\varphi,E}}\simeq \frac{R_0/L_T}{\Omega_{\varphi,d} R_0/v_{ti}}\frac{1}{Pr_E}\left(\frac{ \Pi_{int}^{\Delta  P_{\varphi}}}{Q_i}\frac{v_{ti}}{2R_0}\right)\frac{1}{R_0} \simeq \frac{0.3}{R_0},
 \end{eqnarray}
where we used the Prandtl number $Pr_E={\chi_{\varphi,E}}/{\chi_i}=0.54$, which is obtained for the case $(\partial {\Omega_{\varphi,E}}/\partial r)(R_0a/ v_{ti})=0.152 $. The time averaged momentum flux due to the different pinches $\Pi_{{int}}^{\Delta  P_{\varphi}}$ is obtained from a GS2 simulation with ${\Omega_{\varphi,d} R_0}/{v_{ti}}=0.087$ and ${\Omega_{\varphi,E} R_0}/{v_{ti}}=-0.087$ while eliminating contributions of other neoclassical effects (i.e. $f^{bg,d}_{1,i}=f^{bg,V_\|}_{1,i}$ and $\Delta \Omega_{\varphi, T}=\partial \Omega_{\varphi,d}/ \partial r=\partial\Omega_{\varphi,E}/\partial r=\phi_{1}^{bg}=0$). The small size of the neoclassical particle flow for the Cyclone Base Case results in the relatively small momentum flux due to the different pinches, compared to the case in \cite{Lee:NF2014}.

The origin of the different pinches is the acceleration term that depends on the type of rotation, $({Z_ie}/{m_ic})\Omega_{\varphi,E}(\mathbf{v}_M \cdot\nabla \psi)({\partial f_i^{tb}}/{\partial \mathcal{E}})$ in (\ref{GKE2}). In the presence of E$\times$B flow, the particle is accelerated within the radial electric field, while the diamagnetic flow does not give acceleration. The acceleration term breaks the symmetry of the turbulence because it is odd in the poloidal coordinate ($\theta$). For example, for a circular tokamak, the term is proportional to $\sin\theta$. 

The different momentum diffusivities result in the piece of the intrinsic momentum transport $\Pi_{{int}}^{\Delta  \chi_{\varphi}}$, shown in Table 2-(b). The size of the momentum flux is proportional to the size of the velocity shear, which is determined by the second derivative of the pressure and temperature in radius. We set the size of the velocity shear  $({\partial \Omega_{\varphi,d}}/{\partial r}) ({R_0a}/{v_{ti}})=-0.152$ by setting  $d(R_0/L_n)/dr=0$ and $d(R_0/L_T)/dr=0$ as done in \cite{Barnes:PRL2013}. The different momentum diffusivities are evaluated by comparing the Prandtl number for the diamagnetic flow and E $\times$ B flow, 
\begin{eqnarray}
\Delta Pr\equiv \frac{\chi_{\varphi,d}-\chi_{\varphi,E}}{\chi_i}\simeq \frac{(R_0/L_T) }{(\partial \Omega_{\varphi,d}/\partial r) (R_0 a/v_{ti})}\frac{a}{R_0}\left(\frac{ \Pi_{int}^{\Delta \chi_\varphi}}{Q_i}\frac{v_{ti}}{2R_0}\right)\simeq 0.33,
 \end{eqnarray}
 where $\Pi_{int}^{\Delta \chi_\varphi}$ is obtained from a GS2 simulation with  $({\partial \Omega_{\varphi,d}}/{\partial r}) ({R_0a}/{v_{ti}})=-0.152$ and  $({\partial \Omega_{\varphi,E}}/{\partial r}) ({R_0a}/{v_{ti}})=0.152$ while eliminating contributions of other neoclassical effects  (i.e. $f^{bg,d}_{1,i}=f^{bg,V_\|}_{1,i}$ and $\Delta \Omega_{\varphi, T}=\Omega_{\varphi,d}=\Omega_{\varphi,E}=\phi_{1}^{bg}=0$).   Because the Prandtl number for the E$\times$B flow is about 0.54, this difference is significant, about $61 \%$. 
  
The origin of the different diffusivities is the other rotation type dependent term, $({1}/{B}) (\psi-\psi_0) (\partial \Omega_{\varphi,E}/\partial \psi)\nabla\psi\times \mathbf{\hat{b}} \cdot  \nabla f_i^{tb}$, in equation (\ref{GKE2}). Generally, the radial shear of the rotation reduces the radial correlation length of the turbulent eddies because an eddy radially aligned at an initial time becomes misaligned in time due to the radial gradient of the toroidal flows \cite{Barnes:PRL2011shear, Highcock:PRL2010}. The rotation type dependent term contributes to the radial misalignment differently for different types of rotation, because the particle orbits are modified not by the pressure gradient but by the radial electric field. We found that the difference of the Prandtl number becomes negligible ($\Delta Pr\simeq 0.003$), if the $E \times B$ shear term on the left hand side, $({1}/{B}) (\psi-\psi_0) (\partial \Omega_{\varphi,E}/\partial \psi)\nabla\psi\times \mathbf{\hat{b}} \cdot  \nabla f_i^{tb}$, is ignored but the velocity shear terms on the right hand side, $({c}/{B}) (\nabla\langle\phi^{tb} \rangle \times \mathbf{\hat{b}} \cdot  \nabla \psi) ({m_iIv_\|^\prime}/{T_iB})({\partial \Omega_{\varphi,E}}/{\partial \psi })f_{i}^{bg(R)}$ for E$\times$B flow  and $({c}/{B}) (\nabla\langle\phi^{tb} \rangle \times \mathbf{\hat{b}} \cdot  \nabla f_{1,i}^{bg,d})$ for the diamagnetic flow, are kept in GS2 simulations.

The difference in the momentum diffusivities depends significantly on the turbulent characteristics. For the plasma parameters in \cite{Lee:NF2014}, the effect is negligible, which results in the small difference of the momentum diffusivities (i.e.$\Delta Pr/ Pr <0.01$).

 The time averaged value of momentum flux due to a poloidal dependence $\Pi_{{int}}^{\theta}$ in Table 2-(c) is obtained from a GS2 simulation with the non-zero $\Delta \Omega_{\varphi, T}$ of the blue graph in figure 3-(b) while eliminating contributions of other neoclassical effects (i.e. $f^{bg,d}_{1,i}=f^{bg,V_\|}_{1,i}$ and $\Omega_{\varphi,d}=\Omega_{\varphi,E}=\partial \Omega_{\varphi,d}/ \partial r=\partial\Omega_{\varphi,E}/\partial r=\phi_{1}^{bg}=0$).  This intrinsic momentum transport $\Pi_{{int}}^{\theta}$ occurs because the temperature driven diamagnetic flow  $\Omega_{\varphi,T}(r,\theta)$ has a poloidal dependence different from that of the E$\times$B flow $\Omega_{\varphi,E}$. According to the definition of $\Delta\Omega_{\varphi,T}(r,\theta)$ in equation (\ref{defdelOmegaT}), there is no net toroidal angular momentum due to  $\Delta\Omega_{\varphi,T}(r,\theta)$ on a flux surface. As shown in the blue graph of figure 3-(b), $\Delta \Omega_{\varphi,T}>0$ at the outer-midplane $(|\theta|\lesssim \pi)$, and $\Delta \Omega_{\varphi,T}<0$ at the inner-midplane $(|\theta|\gtrsim \pi)$. The rotation $\Delta \Omega_{\varphi,T}$ can result in a finite flux of toroidal angular momentum because the $E\times B$ radial drift due to the turbulence is nonuniform poloidally. Usually, the $E\times B$ radial drift is larger at the outer-midplane than at the inner-midplane. Thus, there is more negative momentum pinch of positive momentum at the outer-midplane ($-P_\varphi \Delta \Omega_{\varphi,T}  <0$ at $\theta=0$) than positive momentum pinch of the negative momentum at the inner-midplane ($-P_\varphi \Delta \Omega_{\varphi,T} >0$ at $\theta=\pi$). On the other hand, there is more positive momentum diffusion due to negative radial gradient of the positive momentum at the outer-midplane ($-\chi_\varphi (\partial \Delta \Omega_{\varphi,T}/\partial r)  >0$ at $\theta=0$) than negative momentum diffusion due to the positive radial gradient of the negative momentum at the inner-midplane ($-\chi_\varphi (\partial \Delta \Omega_{\varphi,T}/\partial r) <0$ at $\theta=\pi$). Thus, the sign and size of $\Pi_{{int}}^{\theta}$ is determined by the two competing terms on the right hand side of (\ref{Pitheta}).

\begin{figure*} [H]
(a)\\
\includegraphics[scale=0.5]{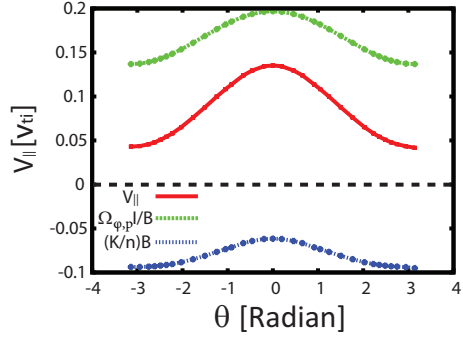}\\
(b)\\
\includegraphics[scale=0.5]{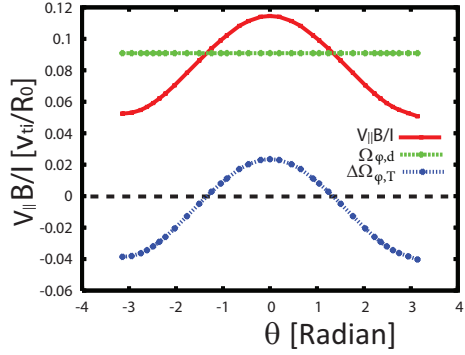}
\caption{The normalized diamagnetic particle flow in terms of poloidal angle calculated by NEO for the Cyclone Base Case with $\rho_\star =0.01$ and low collisionality $\nu_\star =0.043$: (a) The parallel particle flow (red) $V_{\|,d}=(\Omega_{\varphi,p}+\Omega_{\varphi,T})I/B$, the pressure-driven flow (green) $\Omega_{\varphi,p}I/B$ and the temperature-driven flow (blue) $\Omega_{\varphi,T}I/B=(K/n)B$, (b) The parallel particle flow divided by $I/B$ (red) $V_{\|,d}(B/I)=\Omega_{\varphi,d}+\Delta \Omega_{\varphi,T}(\theta)$, the poloidally uniform angular frequency piece (green) $\Omega_{\varphi,d}=(\Omega_{\varphi,p}+\overline{\Omega_{\varphi,T}})
$, and the poloidally non-uniform angular frequency piece (blue) $\Delta \Omega_{\varphi,T}(\theta))$.  Here, $\Omega_{\varphi,p}R_0/ v_{ti}=0.167 $ and $(K/n)B_0/v_{ti}\simeq-0.080$ result in ${\Omega_{\varphi,d} R_0}/{v_{ti}}=0.087$}.
\label{fig:rotation_dia}
\end{figure*}

\Table{\label{tab:pinch_diff}The normalized intrinsic momentum flux caused by the diamagnetic particle flow $\Pi_{{int}}^{V_\|}$ in GS2 for the Cyclone Base Case with $\rho_\star =0.01$,  $d(R_0/L_n)/dr=0$, $d(R_0/L_T)/dr=0$ and low collsionality $\nu_\star =0.043$. Here, $\Pi_{{int}}^{V_\|}$ has three contributions due to : (a) the different momentum pinches $\Pi_{{int}}^{\Delta  P_{\varphi}}$, (b) the different momentum diffusivities $\Pi_{{int}}^{\Delta  \chi_{\varphi}}$ and (c) the different poloidal dependence of the temperature and pressure gradient driven flows $\Pi_{{int}}^{\theta}$.}
\br
&\centre{3}{Characteristics of non-Maxwellian $f_{1,i}^{bg,d}$}& \centre{2}{GS2 results} \\
& $\frac{\Omega_{\varphi,d} R_0}{v_{ti}}$ & $\frac{\partial \Omega_{\varphi,d}}{\partial r} \frac{R_0a}{v_{ti}}$ &$\Delta \Omega_{\varphi,T}(\psi,\theta)$   &$\frac{\Pi_{int}}{Q_i}\frac{v_{ti}}{R_0}$ &$\sigma\left(\frac{ \Pi_{int}}{Q_i}\frac{v_{ti}}{R_0}\right)$\\
\mr
(a) $\Pi_{{int}}^{\Delta  P_{\varphi}}$&  0.087 & 0.0 & 0.0  & -0.005 & 0.019\\
(b) $\Pi_{{int}}^{\Delta  \chi_{\varphi}}$&  0.0 & -0.152 & 0.0 &0.043  & 0.013\\
(c) $\Pi_{{int}}^{\theta}$&  0.0 & 0.0 & Fiq. 3-(b) blue graph & -0.012 & 0.015\\
\br
\end{tabular}
\end{indented}
\end{table}

\clearpage


\subsubsection{Neoclassical parallel heat flow contribution.}\label{heat_flow}

The neoclassical parallel heat flow results in momentum transport $\Pi_{{int}}^{q_\|}$ because it breaks the symmetry of the turbulence even in the absence of parallel momentum (i.e. $\int dv_\| v_\| f_{1,i}^{bg, q_\|} =0$). 
The piece for the parallel heat flow $f^{bg,q_\|}_{1,i}$ is obtained by solving (\ref{DKE0}) with NEO. The normalized size of the heat flow $q_\|/v_{ti}p_i$ is comparable to the normalized size of the particle flow $V_{\|,d}/v_{ti}$, as shown in figure 2. The size of the heat flow is proportional to the temperature gradient and depends on collisionality, as the temperature gradient driven diamagnetic particle flow ($\Omega_{\varphi,T}$) does. We found that the momentum transport due to the heat flow is significant, as shown in Table 1. To calculate the time averaged value of momentum flux due to the parallel heat flow $\Pi_{{int}}^{q_\|}$, we use a GS2 simulation with non-zero $q_\|$ while eliminating contributions of other neoclassical effects  (i.e. $f^{bg,d}_{1,i}=f^{bg,q_\|}_{1,i}$ and $\phi_{1}^{bg}=0$).  For the cyclone case with low collisionality, $\Pi_{{int}}^{q_\|}$ is negative and it dominates over $\Pi_{{int}}^{V_\|}$, giving the negative total momentum flux $\Pi_{{int}}$. 

\begin{figure}
(a)
\\ \includegraphics[scale=0.4]{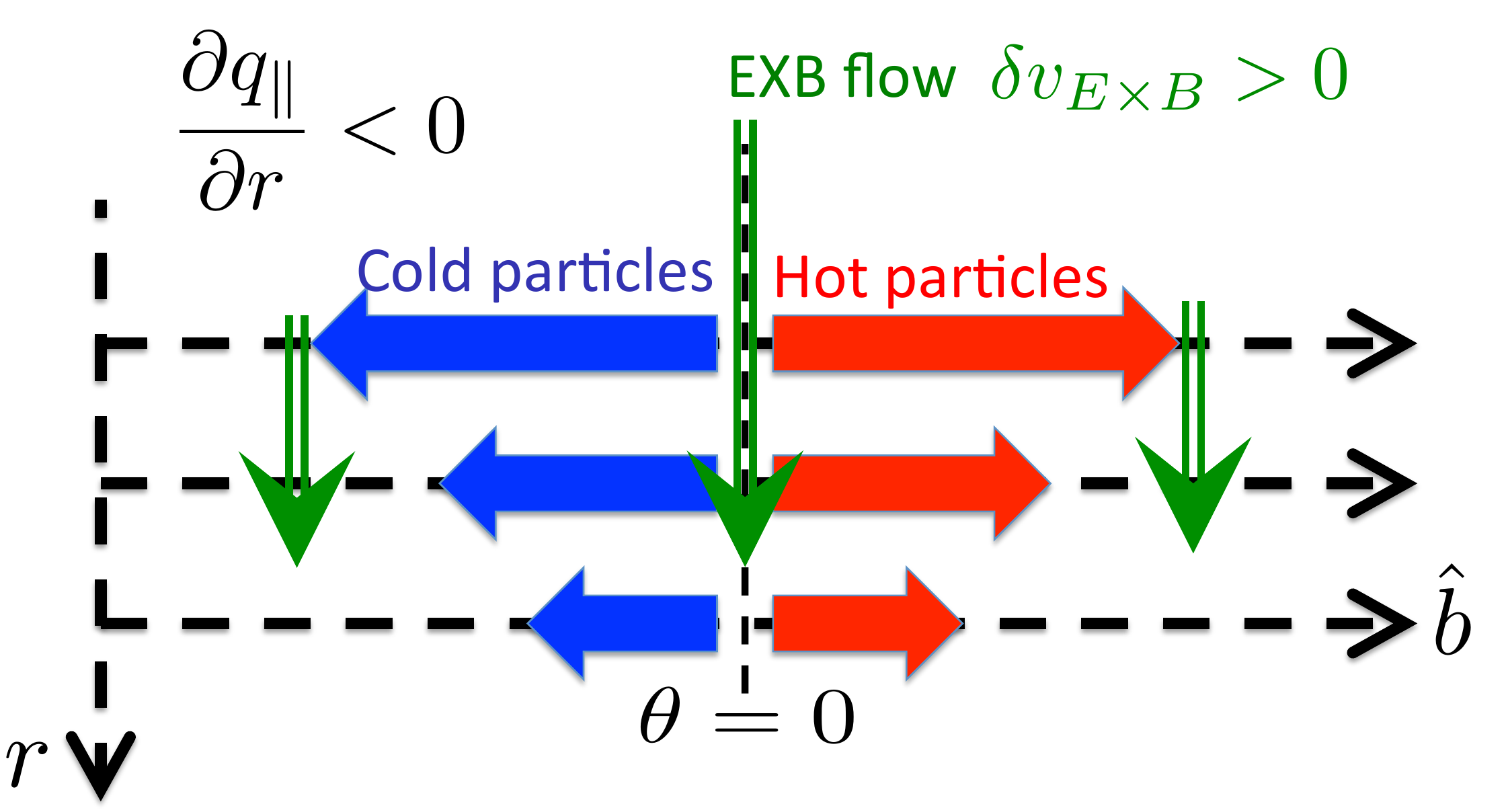}\\
(b)\\ \includegraphics[scale=0.4]{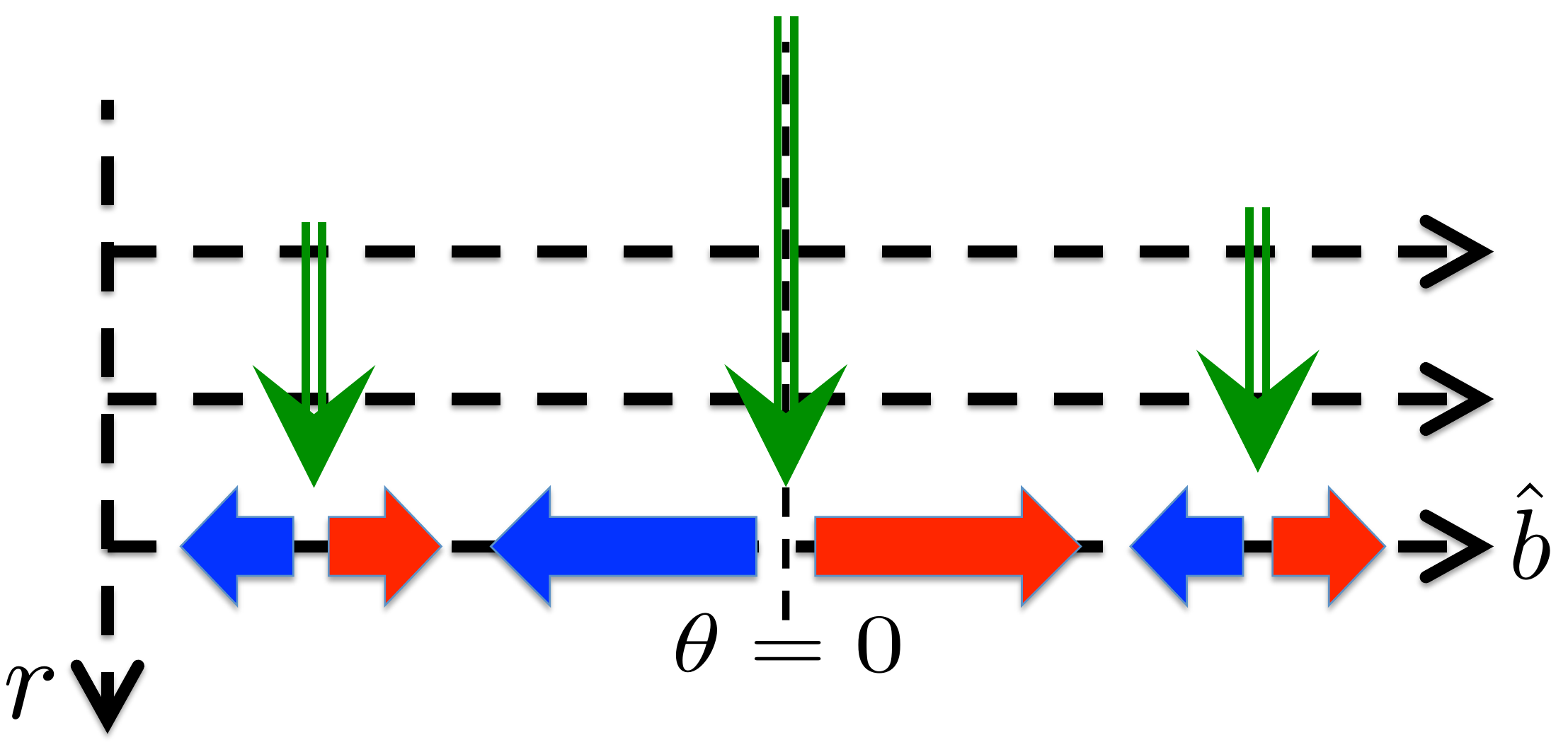}\\
(c)\\ \;\;\;\;\includegraphics[scale=0.4]{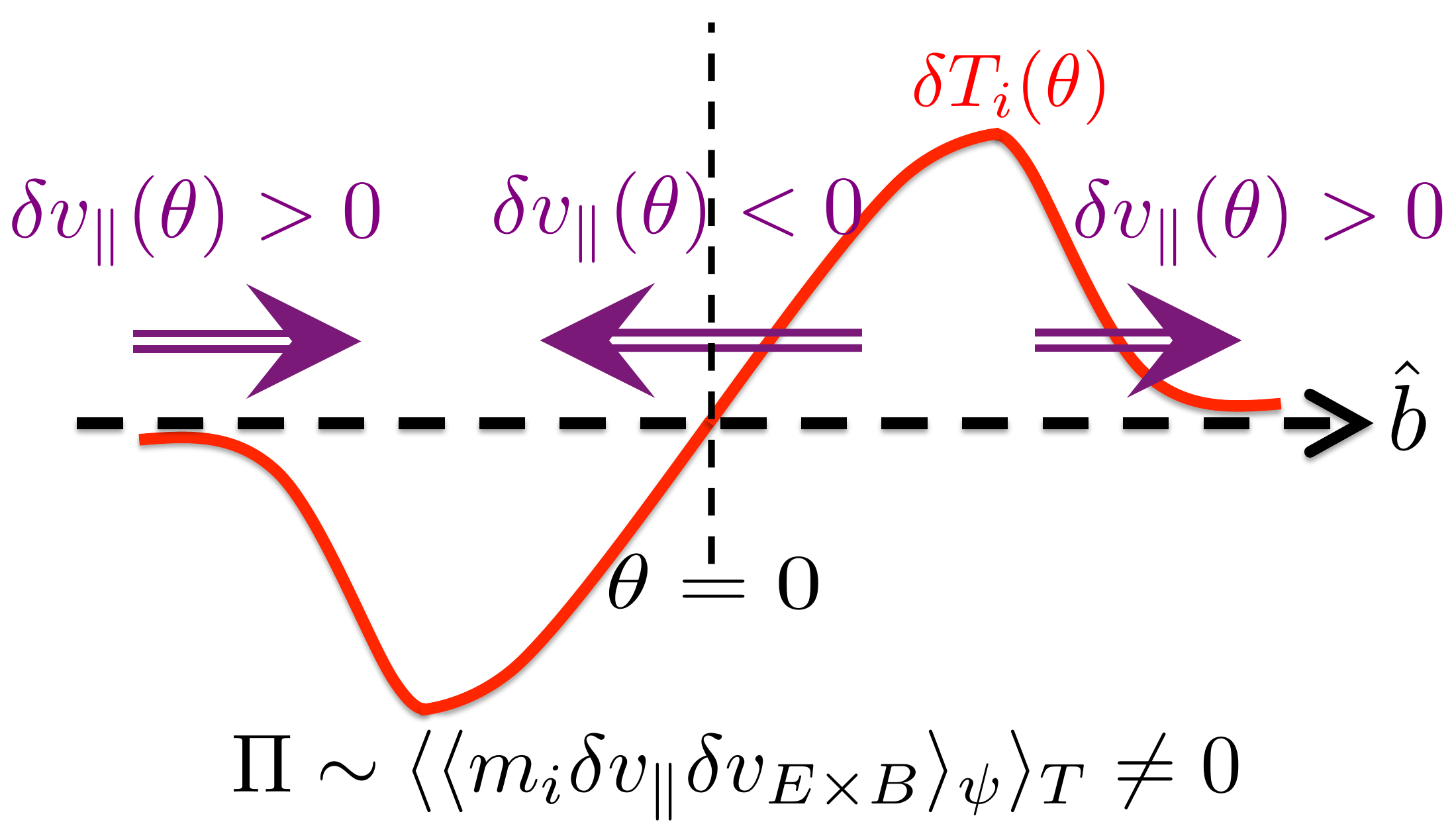}
\caption{A sketch of a the mechanism that leads to momentum transport when there is a radial gradient of neoclassical parallel heat flow, $\Pi_{{int}}^{q_\|}$. (a) Assume that the neoclassical heat flow at inner radii is larger than the heat flow at outer radii. A positive heat flow without parallel particle flow means that hot particles move in the positive parallel direction and cold particles move in the negative parallel direction with the equal speeds. (b) The E $\times$ B flow $\delta v_{E\times B}$, which is larger at the outer-midplane around $\theta=0$ (ballooning structure), gives a poloidally varying perturbed heat flow $\delta q_\|$. (c) The variation of the temperature in the parallel direction $\delta T_i(\theta)$ due to the polloidally varying $\delta q_\|$ results in the poloidally varying parallel particle flow $\delta v_\|(\theta)$, giving non-zero momentum flux as $\delta v_\|(\theta)$ is transported by the radial E $\times$ B drift $\delta v_{E\times B}$.}\label{fig:qpar_mech}
\end{figure}

Figure \ref{fig:qpar_mech} explains the physical origin of the momentum transport due to the radial variation of the parallel heat flow. Because the turbulent radial E $\times$ B flow $\delta v_{E\times B}$ has poloidal variation, the radial change of the parallel heat flow results in a temperature perturbation with poloidal dependence as shown in figure \ref{fig:qpar_mech}-(b). The temperature perturbation can be approximately described by 
\begin{eqnarray}
\delta T_i \sim \frac{\tau_{nl}}{R_0 q}\frac{\partial}{\partial \theta}\left(\delta v_{E\times B}  \tau_{nl}\frac{\partial q_\|}{\partial r} \right),
\end{eqnarray}
where $\tau_{nl}$ is the turbulent nonlinear correlation time.
 The pressure gradient in the parallel coordinate between the higher temperature location and the lower temperature location gives the perturbed parallel particle flow 
 \begin{eqnarray}
m_i\delta v_\| \sim {\tau_{nl}}\frac{n_i}{R_0 q}\frac{\partial \delta T_i}{\partial \theta},
\end{eqnarray}
where the momentum equation in the parallel direction has been used.
The product of perturbed parallel particle flow induced by the heat flow gradient and the turbulent radial E $\times$ B flow $\delta v_{E\times B}$ does not necessarily vanish in the poloidal angle integration, and it results in non-zero momentum flux $\Pi_{{int}}^{q_\|}\sim \langle\langle m_i \delta v_\| \delta v_{E\times B} \rangle_\psi\rangle_T$, as shown in figure \ref{fig:qpar_mech}-(c). 
This mechanism gives $\Pi_{{int}}^{q_\|} < 0$ for $\partial q_\|/\partial r <0$, if $\delta v_\|$ at large poloidal angle is sufficiently small, which is consistent with the results in Table 3.


Table 3 shows the simulation results of intrinsic momentum flux due to the parallel heat flows that are poloidally uniform to remove the poloidal variation effect of the heat flows for simplicity. Significant momentum flux occurs due to the radial gradient of the heat flow, and the sign of the flux in Table 3-(c) is consistent with the mechanism in figure \ref{fig:qpar_mech}, explaining the result in Table 1-(c). In the absence of the radial variation and poloidal variation of the heat flow, the momentum flux is small, as shown in Table 3-(a). 


\Table{Normalized intrinsic momentum flux caused by the diamagnetic parallel heat flow in GS2. The heat flow is similar in size to the heat flow for the Cyclone Base Case in figure 2, but the poloidal variation is not considered for simplicity. We consider the heat flow with (a) no radial gradient, (b) negative gradient and (c) positive gradient.}
\br
&\centre{2}{The input of $f_{1,i}^{bg,q_\|}$}& \centre{2}{GS2 results} \\
& $q_\|/v_{ti}p_i$ & $(\partial q_\|/\partial r)(a/v_{ti}p_i)$&$\frac{\Pi_{int}}{Q_i}\frac{v_{ti}}{R_0}$&$\sigma\left(\frac{ \Pi_{int}}{Q_i}\frac{v_{ti}}{R_0}\right)$\\
\mr
(a) Constant $q_\|$ & 0.1 & 0  & -0.005 & 0.016\\
(b) Positive $(\partial q_\|/\partial r)$&  0& 0.5 &0.057  & 0.018\\
(c) Negative $(\partial q_\|/\partial r)$& 0 & -0.5  & -0.067 & 0.018\\
\br
\end{tabular}
\end{indented}
\end{table}

\subsection{Moderate collisionality}\label{sec:highcol}

Table \ref{tab:hichcol1} shows GS2 results for the contribution of each neoclassical correction to the momentum flux for moderate collisionality $\nu_\star \simeq 1$. In the moderate collisionality case, the momentum flux due to the neoclassical potential $\phi_1^{bg}$ and the distribution function $f^{bg,other}_{1,i}$ are non-negligible. The different results in Table 1 and \ref{tab:hichcol1} can be used to explain why the total momentum flux is reversed as the collisionality increases, as found in \cite{Barnes:PRL2013}. For the Cyclone Base Case, as the collisionality increases ($\nu_\star =0.043 \rightarrow 4.3$), the momentum flux changes its sign from negative to positive ($(\Pi_{int}/{Q_i})({v_{ti}}/{R_0})=-0.022 \rightarrow 0.049$). The sign reversal is due to several combined effects. 

One is the increase of the positive momentum flux due to the increase of the diamagnetic particle flow. Because strong collisions reduce the temperature driven diamagnetic flow, whose direction is negative in this case, the total size of the diamagnetic flow and its radial shear increases ($(\Omega_{\varphi,d}R_0/v_{ti}) =0.087 \rightarrow 0.180$ and $d(\Omega_{\varphi,d}R_0/v_{ti})/dr =-0.152 \rightarrow -0.279$). Compare the red graphs in figure 5-(a) and figure 2-(a). For this reason, the positive momentum flux $\Pi_{{int}}^{\Delta  P_{\varphi}}+\Pi_{{int}}^{\Delta  \chi_{\varphi}}$ increases, as shown in Table \ref{tab:highcol2}-(a) and (b) compared with Table 2-(a) and (b). The second effect is the reduced negative momentum flux due to the reduced poloidal dependence effect ($(\Pi_{int}^{\theta}/{Q_i})({v_{ti}}/{R_0})=-0.012$ in Table 2-(c) $\rightarrow 0.001$ in Table \ref{tab:highcol2}-(c)). Compare the blue graphs in figure 5-(b) and  figure 3-(b). This reduction is also due to the reduced temperature driven diamagnetic flow. The third effect is the reduced negative momentum flux due to the heat flow ($(\Pi_{int}^{q_\|}/{Q_i})({v_{ti}}/{R_0})=-0.075 \rightarrow -0.045$), which is due to the reduced size of the heat flow. Compare the green graphs in figure 5-(a) an figure 2-(a). There is also a contribution due to $\phi_{1}^{bg}$ and $f^{bg,other}_{1,i}$ for moderate collisionality, giving the positive momentum flux $\Pi_{{int}}^{\phi_1^{bg}} + \Pi_{{int}}^{other}$ in Table \ref{tab:hichcol1}.

\begin{figure*} 
(a)
\\ \includegraphics[scale=0.5]{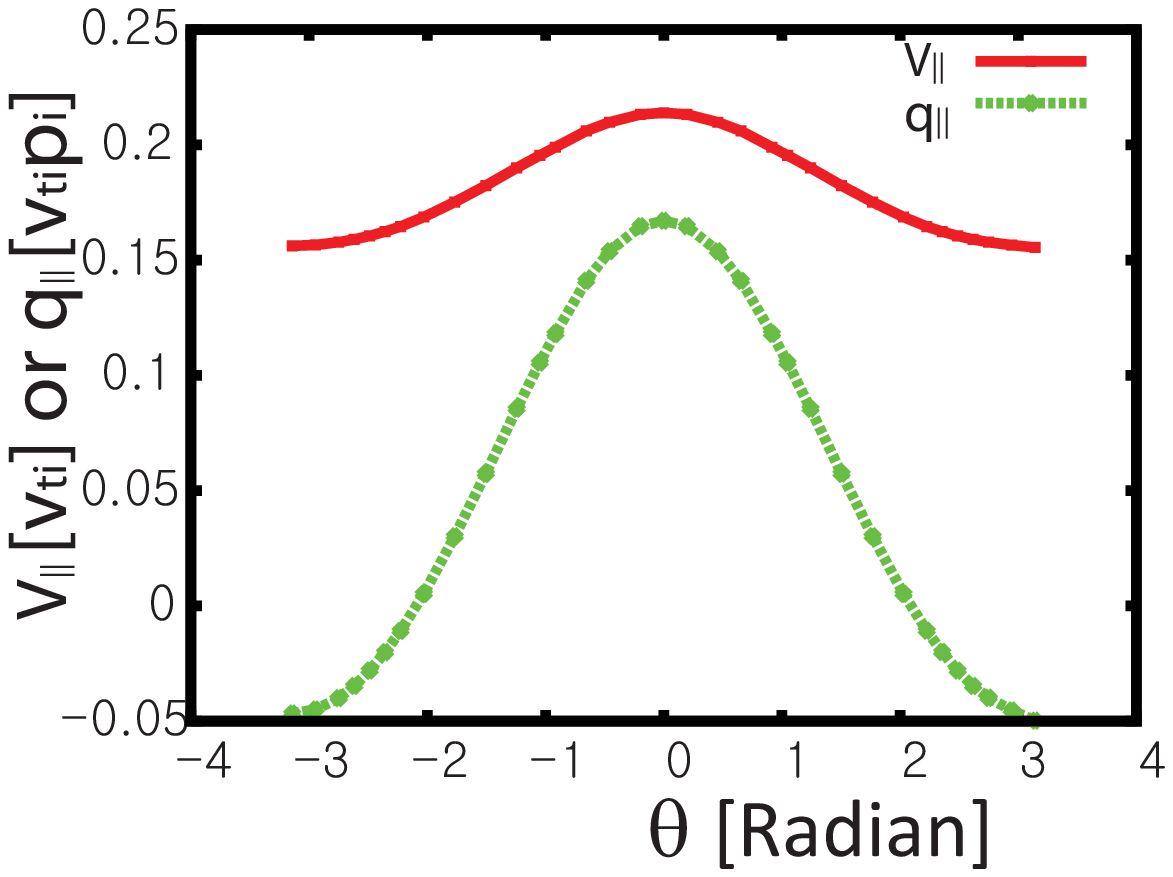} \\
(b) \\ 
\includegraphics[scale=0.5]{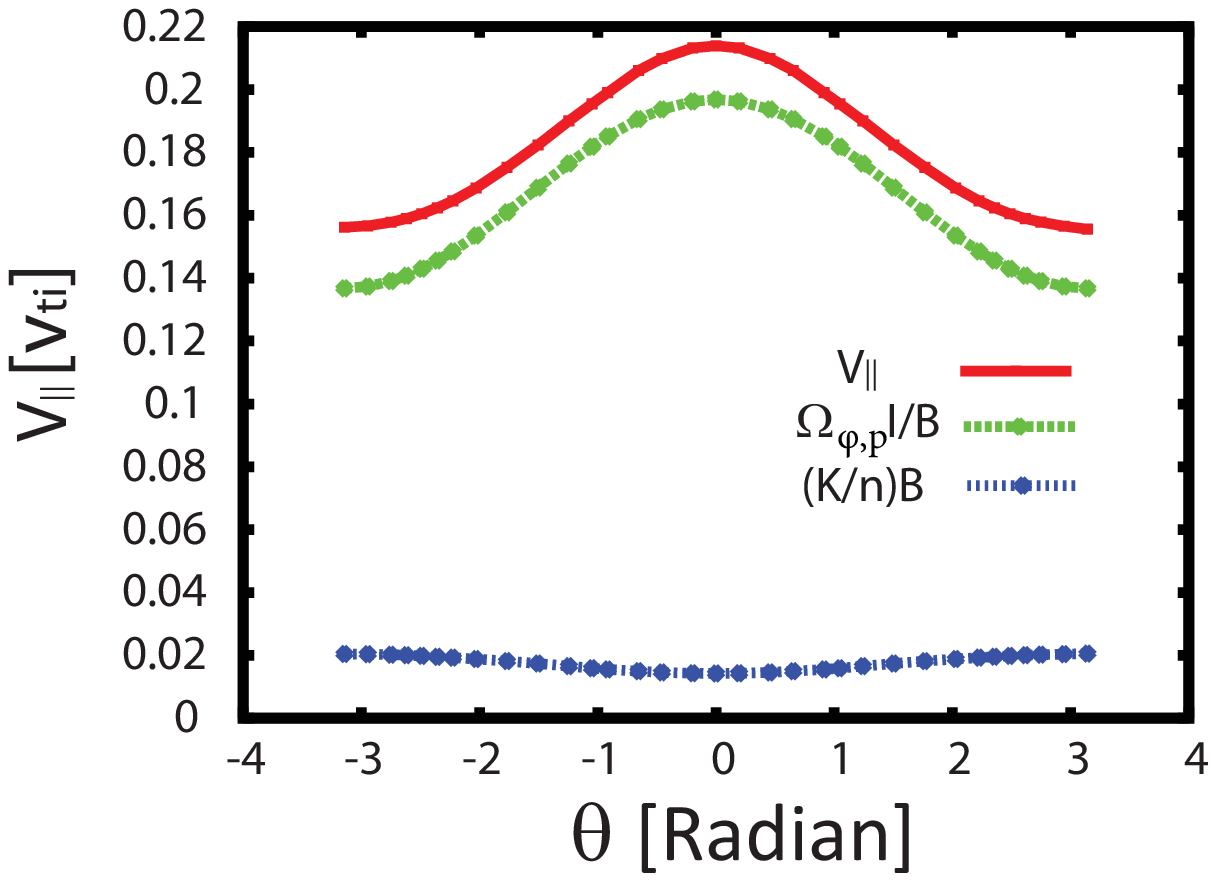}
\caption{The normalized diamagnetic particle flow in terms of poloidal angle calculated by NEO for the Cyclone Base Case with $\rho_\star =0.01$ and moderate collisionality $\nu_\star =4.3$. (a)  The parallel particle flow (red) $V_{\|,d}/v_{ti}$ and the parallel heat flow $q_\|/v_{ti}p_i$ (green). (b) the parallel particle flow (red) $V_{\|,d}=(\Omega_{\varphi,p}+\Omega_{\varphi,T})I/B$, the pressure-driven flow (green) $\Omega_{\varphi,p}I/B$ and the temperature-driven flow (blue) $\Omega_{\varphi,T}I/B=(K/n)B$. Here, $\Omega_{\varphi,p}R_0/ v_{ti}=0.167 $ and $(K/n)B_0/v_{ti}\simeq 0.013$.
}.
\label{fig:rotation_dia}
\end{figure*}

\Table{\label{tab:hichcol1}The radial flux of the toroidal angular momentum $\Pi_{int}$ in the non-rotating state ($\Omega_{\varphi}=\Omega_{\varphi,d} +\Omega_{\varphi,E}=0$), divided by the ion heat flux $Q_i$ given by GS2 for the Cyclone Base Case with $\rho_\star =0.01$, $d(R_0/L_n)/dr=0$, $d(R_0/L_T)/dr=0$ and moderate collisionality $\nu_\star =4.3$  for (a) full neoclassical function and potential, (b) only the particle flow piece $f^{bg,V_\|}_{1,i}$, (c) only the parallel heat flow piece $f_{1,i}^{bg,q_\|}$, (d) only the neoclassical potential and (e) only the remaining piece $f_{1,i}^{bg,other}$. Here, $\sigma\left(\frac{ \Pi_{int}}{Q_i}\frac{v_{ti}}{R_0}\right)$ is the standard deviation of the normalized momentum flux from the time averaged value due to the turbulent fluctuations.}
\br
&\centre{1}{Size}& \centre{2}{GS2 results} \\
&  &$\frac{\Pi_{int}}{Q_i}\frac{v_{ti}}{R_0}$ &$\sigma\left(\frac{ \Pi_{int}}{Q_i}\frac{v_{ti}}{R_0}\right)$\\
\mr
(a) $\Pi_{{int}}$  & $f^{bg,V_\|}_{1,i}+f^{bg,q_\|}_{1,i}+f^{bg,other}_{1,i}$  & 0.049 & 0.035 \\
(b) $\Pi_{{int}}^{V_\|}$ & $V_{\|,d}$ in figure 5-(a) & 0.058 & 0.027\\
(c) $\Pi_{{int}}^{q_\|}$  & $q_\|$ in figure 5-(a)  &-0.046 & 0.033\\
(d) $\Pi_{{int}}^{\phi_{1}^{bg}}$ &$e\sqrt{\langle (\phi_{1}^{bg})^2 \rangle_\psi}/T_i \simeq$ 0.009  &-0.011 & 0.036 \\
(e) $\Pi_{{int}}^{other}$ & other than $V_{\|,d}$ and $q_\|$&0.034 & 0.028 \\
\br
\end{tabular}
\end{indented}
\end{table}

\Table{\label{tab:highcol2}The normalized intrinsic momentum flux caused by the diamagnetic particle flow $\Pi_{{int}}^{V_\|}$ in GS2 for the Cyclone Base Case with $\rho_\star =0.01$, $d(R_0/L_n)/dr=0$, $d(R_0/L_T)/dr=0$ and moderate collisionality $\nu_\star =4.3$. Here, $\Pi_{{int}}^{V_\|}$ has three contributions from (a) the different momentum pinches $\Pi_{{int}}^{\Delta  P_{\varphi}}$, (b) the different momentum diffusivities $\Pi_{{int}}^{\Delta  \chi_{\varphi}}$ and (c) the different dependence on the poloidal angle $\Pi_{{int}}^{\theta}$.}
\br
&\centre{3}{Characteristics of non-Maxwellian $f_{1,i}^{bg,d}$}& \centre{2}{GS2 results} \\
& $\frac{\Omega_{\varphi,d} R_0}{v_{ti}}$ & $\frac{\partial \Omega_{\varphi,d}}{\partial r} \frac{R_0a}{v_{ti}}$ &$\Delta \Omega_{\varphi,T}(\psi,\theta)$   &$\frac{\Pi_{int}}{Q_i}\frac{v_{ti}}{R_0}$ &$\sigma\left(\frac{ \Pi_{int}}{Q_i}\frac{v_{ti}}{R_0}\right)$\\
\mr
(a) $\Pi_{{int}}^{\Delta  P_{\varphi}}$&  0.180 & 0.0 & 0.0  &-0.024 & 0.022 \\
(b) $\Pi_{{int}}^{\Delta  \chi_{\varphi}}$&  0.0 & -0.279 & 0.0 &0.066  &  0.030\\
(c) $\Pi_{{int}}^{\theta}$&  0.0 & 0.0 & Fiq. 5-(b) blue graph &0.001  &  0.029 \\
\br
\end{tabular}
\end{indented}
\end{table}

\subsubsection{Neoclassical potential.}\label{neo_potential}
 To find the contribution of the neoclassical potential, we calculate the time averaged momentum transport in GS2 simulations by eliminating the neoclassical distribution function $f_{1,i}^{bg,d}=0$ but keeping the potential from NEO $\phi_{1}^{bg}$.  Table \ref{tab:neopot} shows the momentum transport in terms of various collisionalities. The size of the neoclassical potential varies significantly with collisionality \cite{hinton1976theory,Helander:Coll}. As the collisionality increases, the size of the neoclassical potential increases and gives a larger momentum transport. The momentum transport due to the neoclassical potential in Table \ref{tab:neopot} does not increase monotonically with increasing collisionality because the neoclassical potential at different radial and poloidal position has a different dependence on the collisionality. 

The effects of neoclassical potential $\phi_1^{bg}(\psi,\theta)$ on the momentum transport are determined by two terms in the higher order gyrokinetic equation (\ref{GKE2}). One term gives the $E\times B$ drift  of the turbulent fluctuations, $-({c}/{B}) \nabla\phi^{bg}_1 \times \mathbf{\hat{b}} \cdot \nabla f^{tb}_i$. 
 The other term corresponds to the parallel neoclassical electric field changing the energy of the particle, 
 \begin{eqnarray}
 -\frac{Z_ie}{m_i}  v_\| \mathbf{\hat{b}}\cdot\nabla \phi^{bg}_1  \frac{\partial f^{tb}_i}{\partial \mathcal{E}}= -\frac{Z_ie}{m_i}  v_\| \mathbf{\hat{b}}\cdot\nabla\theta \frac{ \partial \phi^{bg}_1}{\partial \theta}  \frac{\partial f^{tb}_i}{\partial \mathcal{E}}\label{neopot2}.
 \end{eqnarray}
 The energy change in the neoclassical potential due to the magnetic drift can be neglected because it is smaller than other terms (i.e. $|\mathbf{v}_M \cdot \nabla \phi^{bg}_1|\ll  |\mathbf{v}_M \cdot \nabla \phi^{bg}_0|$ and $\mathbf{v}_M \cdot \nabla \phi^{bg}_1\ll  v_\| \mathbf{\hat{b}}\cdot\nabla \phi^{bg}_1$). 
 
\Table{\label{tab:neopot}The radial turbulent flux  $\Pi_{{int}}^{\phi_{1}^{bg}}$ in GS2 due to the neoclassical potential in the cyclone case for various collisionalities $\nu_\star$. The NEO code is used to find the normalized neoclassical potentials $e\sqrt{\langle (\phi_{1}^{bg})^2 \rangle_\psi}/T_i $ at $r/a=0.54$. For the results in this table, we assume $d(R_0/L_n)/dr=0$ and $d(R_0/L_T)/dr=0$. The radial gradient of the potential $(ae/T_i) (\partial\sqrt{\langle (\phi_{1}^{bg})^2 \rangle_\psi}/\partial r)$ is obtained from NEO results at two different radius at $r/a=0.53$ and $r/a=0.55$ using finite difference.}
\br
&\centre{2}{NEO results}& \centre{2}{GS2 results} \\
 $\nu_\star$ & $e\sqrt{\langle (\phi_{1}^{bg})^2 \rangle_\psi}/T_i $&  $(ae/T_i) (\partial\sqrt{\langle (\phi_{1}^{bg})^2 \rangle_\psi}/\partial r)$ & $\frac{\Pi_{int}^{\phi_{1}^{bg}}}{Q_i}\frac{v_{ti}}{R_0}$ &$\sigma\left(\frac{ \Pi_{int}^{\phi_{1}^{bg}}}{Q_i}\frac{v_{ti}}{R_0}\right)$\\
\mr
(a) 0.043 & 0.0005& 0.0001&  0.002&  0.019\\
(b) 0.12  &  0.0012& 0.0000& 0.003& 0.020\\
(c) 0.43 & 0.0027&-0.0006&-0.011 & 0.026 \\
(d) 1.2 & 0.0048&-0.0008& -0.015 &0.033 \\
(e) 4.3 &0.0092&0.0070 &-0.011 & 0.036  \\
\br
\end{tabular}
\end{indented}
\end{table}
 
\subsubsection{Other neoclassical contributions.}\label{neo_higher}

 To calculate the neoclassical contributions that are not the neoclassical particle and heat flow and the potential, we use GS2 simulations with $\phi_{1}^{bg}=0$ and $f^{bg,other}_{1,i}$, which is the remaining piece of the original distribution function from NEO once the particle flow and the parallel heat flow pieces are subtracted, $f_{1,i}^{bg}-f^{bg,V_\|}_{1,i}-f^{bg,q_\|}_{1,i}$. Here, $f^{bg,V_\|}_{1,i}$ and $f^{bg,q_\|}_{1,i}$ are reconstructed from equations (\ref{neo2}) and (\ref{neo_qpar}) using $V_{\|,d}$ and $q_\|$ of  $f_{1,i}^{bg}$ from NEO. 
 
As the size of $f^{bg,other}_{1,i}$ increases with collisionality, the contribution $\Pi_{int}^{other}$ generally increases with collisionality. Table 7 shows the momentum transport in terms of various collisionalities. The contribution $\Pi_{int}^{other}$ is approximately zero for low to moderate collisionalities.  At the highest collisionality considered, $\Pi_{int}^{other}$ takes on an appreciable positive value that impacts the overall momentum flux.  This contribution is related to $\Pi_{{int}}^{\phi_{1}^{bg}}$ because $f^{bg,other}_{1,i}$ contains the poloidallly-dependent density and temperature that drive $\phi_1^{bg}$. As shown in section \ref{neo_potential} and Table 6, the contribution $\Pi_{{int}}^{\phi_{1}^{bg}}$ increases with collisionality. 



\Table{\label{tab:neo_higher}The radial turbulent flux  $\Pi_{{int}}^{other}$ in GS2 due to $f^{bg,other}_{1,i}$ for various collisionalities $\nu_\star$ in the same cyclone case as Table \ref{tab:neopot}.}
\br
& \centre{2}{GS2 results} \\
 $\nu_\star$  & $\frac{\Pi_{int}^{other}}{Q_i}\frac{v_{ti}}{R_0}$ & $\sigma\left(\frac{ \Pi_{int}^{other}}{Q_i}\frac{v_{ti}}{R_0}\right)$\\
\mr
(a) 0.043 &   0.001&  0.016\\
(b) 0.12  &  -0.004 & 0.018\\
(c) 0.43 &0.005 & 0.021 \\
(d) 1.2 & -0.006 &0.028 \\
(e) 4.3 &0.034 & 0.028 \\
\br
\end{tabular}
\end{indented}
\end{table}
 
 \section{Discussion} 

In this paper, the turbulent momentum transport in the absence of total toroidal flow is investigated. 
We have studied contributions to the momentum transport due to neoclassical effects. The momentum transport has a piece due to the difference in the momentum pinches $\Pi_{{int}}^{\Delta  P_{\varphi}}$ and a piece due to different diffusivities $\Pi_{{int}}^{\Delta  \chi_{\varphi}}$ between the diamagnetic particle flow and $E\times B$ flow. Even for diamagnetic particle flow that does not have flux surface averaged toroidal angular momentum, the different poloidal dependence between the temperature driven diamagnetic flow and the pressure driven diamagnetic flow can result in the momentum transport $\Pi_{{int}}^{\theta}$. Also, there are contributions due to the neoclassical parallel heat flow $\Pi_{{int}}^{q_\|}$ and the neoclassical potential $\Pi_{{int}}^{\phi_1^{bg}}$. Those contributions are evaluated  for an example case using higher order gyrokinetics corrected to retain the neoclassical flow effects. We have shown that the relative importance of all these contributions changes with collisionality. 

The size and the sign of each contribution evaluated in section \ref{sec:int_mom} for the Cyclone Base Case using a set of plasma parameters may not be generalizable to the whole parameter phase space. For the numerical results in this paper, $d(R_0/L_n)/dr=0$ and $d(R_0/L_T)/dr=0$ are chosen, however changes in these second derivatives can modify the contributions to the momentum transport significantly due to the changes in $dV_{\|,d}/dr$ and $dq_\|/dr$. Nevertheless, clasifying the components of momentum transport by their driving mechanisms as is done in this paper is useful to analyze and interpret momentum transport with other plasma parameters.

\pagebreak


\end{document}